\documentclass[reqno,11pt]{amsart}
\usepackage[utf8]{inputenc}
\usepackage{enumitem}
\usepackage{graphicx}
\usepackage{amscd}
\usepackage{slashed}
\usepackage{amssymb}
\usepackage{mathtools}
\usepackage[mathscr]{eucal}
\usepackage[linktocpage]{hyperref}
\usepackage{pstricks}
\usepackage{pst-all}
\usepackage{pst-all}

\numberwithin{equation}{section}
\allowdisplaybreaks[1]
\definecolor{labelkey}{gray}{.65}

\title[The Isoperimetric Flow for Causal Variational Principles]{The Isoperimetric Flow and Synthetic Scalar Curvature for Causal Variational Principles}

\author[F.\ Finster]{Felix Finster}
\address{Fakult\"at f\"ur Mathematik \\ Universit\"at Regensburg \\ D-93040 Regensburg \\ Germany}
\email{finster@ur.de}

\author[F.\ Gmeineder]{Franz Gmeineder}
\address{Fachbereich Mathematik \\ Technische Universität Darmstadt \\ D-64277
Darmstadt \\ Germany}
\email{gmeineder@mathematik.tu-darmstadt.de}

\author[M.\ Kraus]{Margarita Kraus \\ \\ November 2026}
\address{Fachbereich 08 \\ Johannes-Gutenberg-Univer\-si\-t\"at Mainz \\
D-55099 Mainz \\ Germany }
\email{mkraus@mathematik.uni-mainz.de}

\newtheorem{Def}{Definition}[section]
\newtheorem{Thm}[Def]{Theorem}
\newtheorem{Prp}[Def]{Proposition}
\newtheorem{Lemma}[Def]{Lemma}
\newtheorem{Remark}[Def]{Remark}

\newcommand{\beq}{\begin{equation}}
\newcommand{\eeq}{\end{equation}}
\newcommand{\Proof}{\begin{proof}}
\newcommand{\QED}{\end{proof} \noindent}
\newcommand{\QEDrem}{\ \hfill $\Diamond$}

\newcommand{\la}{\langle}
\newcommand{\ra}{\rangle}

\newcommand{\R}{\mathbb{R}}
\newcommand{\1}{\mbox{\rm 1 \hspace{-1.05 em} 1}}

\newcommand{\N}{\mathbb{N}}

\newcommand{\G}{\mathscr{G}}

\newcommand{\bep}{\begin{pmatrix}}
\newcommand{\enp}{\end{pmatrix}}

\renewcommand{\O}{\mathscr{O}}

\renewcommand{\O}{{\mathscr{O}}}
\renewcommand{\L}{{\mathcal{L}}}
\newcommand{\Sact}{{\mathcal{S}}}
\newcommand{\s}{{\mathfrak{s}}}

\newcommand{\scrM}{\myscr M}
\newcommand{\scrN}{\myscr N}

\renewcommand{\div}{{\rm{div}}\,}
\newcommand{\x}{x}
\newcommand{\y}{y}

\newcommand{\meas}{\mathfrak{M}}

\DeclareFontFamily{OT1}{rsfso}{}
\DeclareFontShape{OT1}{rsfso}{m}{n}{ <-7> rsfso5 <7-10> rsfso7 <10-> rsfso10}{}
\DeclareMathAlphabet{\myscr}{OT1}{rsfso}{m}{n}

\DeclareMathOperator{\Ric}{Ric}

\renewcommand{\Tr}{\text{\rm{Tr}}}

\DeclareMathOperator{\supp}{supp}

\renewcommand{\v}{\mathfrak{v}}

\newcommand{\bitem}{\begin{itemize}[leftmargin=2.5em]}
\newcommand{\eitem}{\end{itemize}}

\begin{document}
\begin{abstract}
An isoperimetric flow is introduced in the setting of causal variational principles.
Describing the whole space by the support of a minimizing measure,
we define an area functional on subsets by a surface layer integral.
Mean curvature is defined via the first variation of this area functional.
Using the method of minimizing movement, the isoperimetric flow is constructed
as a Lipschitz curve of measures with prescribed volume change.
As illustrating examples, we introduce and analyze ``smeared variants'' of the mean curvature
flow in Euclidean space and on a Riemannian manifold.
In these examples, it is shown that our area functional coincides with the usual area
up to specified curvature corrections.
As an application of our general results, by analyzing the volume and area of small isoperimetric balls,
we introduce notions of local dimension and synthetic scalar curvature.
\end{abstract}

\maketitle
\tableofcontents
\section{Introduction} \label{secintro}
\subsection{Aims and Scope}
Geometric notions such as surface area, dimension and curvature are traditionally formulated in terms of an underlying metric structure. In the smooth setting, they are closely tied to the differential geometry of Riemannian manifolds, while various extensions to non-smooth spaces have been developed within metric geometry and geometric measure theory. A different approach is to investigate to what extent geometric quantities can be defined directly in terms of a measure and a prescribed interaction between points, without requiring an underlying metric or differentiable structure.

The present paper develops such an approach in the framework of nonlocal interaction kernels. Starting from a non-negative kernel, we introduce a notion of nonlocal surface area and study the associated isoperimetric evolution. We then investigate how its small-scale behavior can be used to characterize geometric quantities, in particular dimension and scalar curvature. The underlying principle is to quantify the interaction between a measurable set and its complement, thereby obtaining a nonlocal analogue of classical surface area. In contrast to the local perimeter, this quantity incorporates a regularization at the scale determined by the interaction kernel. For suitable kernels on smooth manifolds, the corresponding asymptotic expansions recover classical geometric quantities, with higher-order terms encoding curvature contributions.

Our framework is that of \emph{causal variational principles}, a mathematical
generalization of the causal action principle from the theory of causal fermion
systems~\cite{continuum,jet}. For the physical background and further
mathematical context, see~\cite{review,dice2014,cfs,intro} and the
website~\cite{cfsweblink}. The basic object is a measure which encodes the
underlying space, or spacetime, together with its additional structures.
We focus on the non-dynamical setting in which the measure~$\mu$ describes
space; this arises, for example, from static causal fermion
systems~\cite[Section~3]{pmt}.
In this situation, the causal action can be written as
\[ 
\Sact (\mu) = \int_\G d\mu(\x) \int_\G d\mu(\y)\: \L(\x,\y) \:, \]
where~$\G$ is a smooth manifold, the {\em{Lagrangian}}~$\L : \G \times \G \rightarrow \R^+_0$ is a given
non-negative function, and~$\mu$ is a Borel measure on~$\G$
(details on the needed regularity assumptions and the corresponding
causal variational principle will be recalled in Section~\ref{secprelim}).

Coming back to the above general statement that the measure also describes all the structures in space
or spacetime, the more specific question arises how and in which sense the measure~$\mu$ encodes
{\em{geometric structures}}, and how these structures are related to standard differential geometric notions.
The present paper is devoted to a first study of these questions
(another, complementary approach has been pursued in~\cite{lcalc, gauss}).
What makes these questions particularly interesting is the fact that the setting of causal variational principles
does not require any smoothness assumptions. In particular, space~$N$ as defined as the
support of the measure of~$\mu$,
\[ N:= \supp \mu \:, \]
does not need to have a smooth manifold structure. Therefore, answering the above questions in the
general setting of causal variational principles gives us at the same time a method for introducing
and analyzing {\em{non-smooth geometries}}.

Having no metric to our disposal, all geometric objects must be formulated purely in terms of
the measure~$\mu$ and the Lagrangian~$\L$.
Our general strategy is to generalize ideas and notions from the theory of geometric flows to our setting.
Due to the inherent non-smoothness, we use De Giorgi's method of {\em{minimizing
movements}}~\cite{degiorgi-minmov, ambrosio-minmov, braides-minmov}
(this method has been used previously for the study of causal variational principles in~\cite{minmov}).
More specifically, in order to explore the local geometry, we shall set up a flow of
sets~$(\Omega_t)_{t \in \R}$ in~$N$ emanating from a single point of~$N$. This flow, referred to as the
{\em{isoperimetric flow}} can be regarded as a non-smooth adaptation of the well-known mean curvature flow
(see for example~\cite{mantegazza, ritore-sinestrari} for the general context).
Analyzing area and volume of the resulting sets~$\Omega_t \subset N$, one can get local information
on isoperimetric properties of~$N$, which in turn make it possible to introduce notions of
{\em{dimension}} and {\em{scalar curvature}} of~$N$.

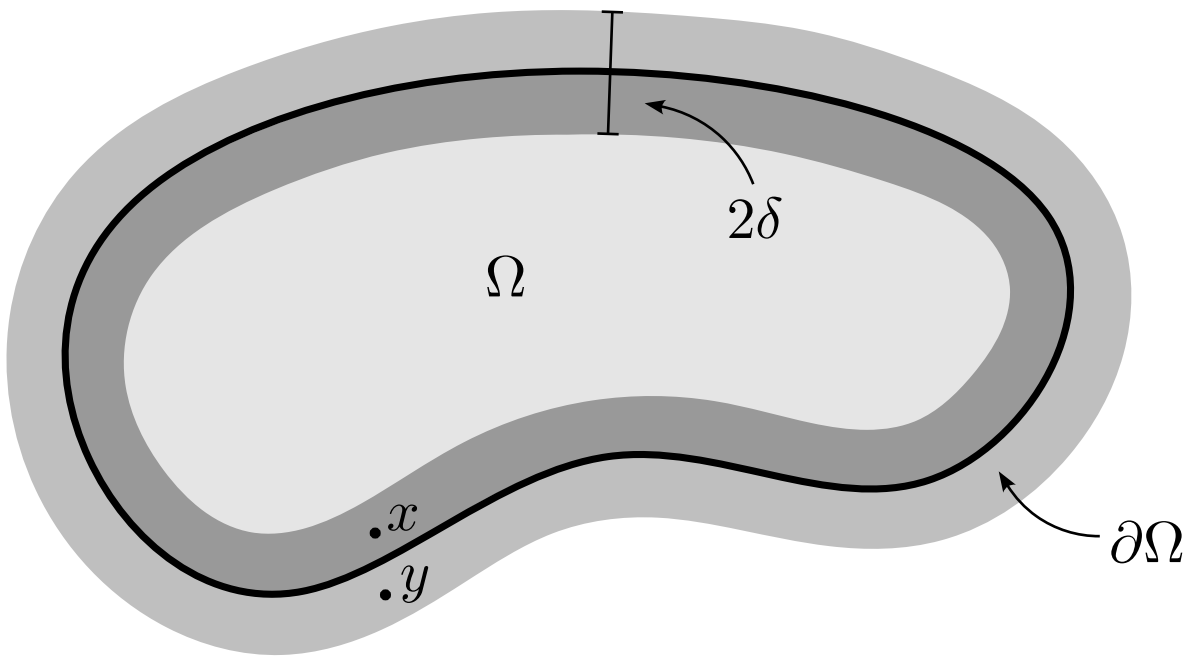
\begin{figure}
\psset{xunit=.5pt,yunit=.5pt,runit=.5pt}
\begin{pspicture}(423.42421644,242.77331255)
{
\newrgbcolor{curcolor}{0.89803922 0.89803922 0.89803922}
\pscustom[linestyle=none,fillstyle=solid,fillcolor=curcolor]
{
\newpath
\moveto(135.03752693,211.62509963)
\lineto(171.74875843,217.85908262)
\lineto(217.4646274,220.62974089)
\lineto(252.79052598,218.89807853)
\lineto(282.22878614,215.08842436)
\lineto(318.59366173,207.122779)
\lineto(340.75892787,200.19612955)
\lineto(364.30954205,189.45982971)
\lineto(383.01147591,176.6455305)
\lineto(395.1331011,158.63625176)
\lineto(399.2891074,146.86094845)
\lineto(399.2891074,130.23699885)
\lineto(396.17213102,116.38371129)
\lineto(387.51380031,99.41342089)
\lineto(377.12384126,85.90646278)
\lineto(364.65586016,75.51649239)
\lineto(351.14888693,67.89718026)
\lineto(333.13962709,63.04853018)
\lineto(307.16471055,63.74119286)
\lineto(284.99944441,68.58984294)
\lineto(255.21484724,75.17016294)
\lineto(228.54725669,75.51648105)
\lineto(207.07465701,69.97516074)
\lineto(185.60204598,61.66318971)
\lineto(165.51477921,49.88789018)
\lineto(144.73483843,38.45892011)
\lineto(127.07188913,29.45427885)
\lineto(112.52592756,24.95196199)
\lineto(98.67262866,23.22029963)
\lineto(85.16567055,24.95196199)
\lineto(74.08303748,29.1079494)
\lineto(59.53707969,37.76624987)
\lineto(46.72278047,48.8488905)
\lineto(33.21581858,69.28249806)
\lineto(24.55751055,90.40876829)
\lineto(21.09418961,113.95937113)
\lineto(25.25018079,139.24162877)
\lineto(36.33281008,159.67524011)
\lineto(49.83976819,173.87486845)
\lineto(69.92705008,187.38182656)
\lineto(93.13131969,199.15712987)
\lineto(115.29658961,207.46909711)
\closepath
}
}
{
\newrgbcolor{curcolor}{0.89803922 0.89803922 0.89803922}
\pscustom[linewidth=0.99999871,linecolor=curcolor]
{
\newpath
\moveto(135.03752693,211.62509963)
\lineto(171.74875843,217.85908262)
\lineto(217.4646274,220.62974089)
\lineto(252.79052598,218.89807853)
\lineto(282.22878614,215.08842436)
\lineto(318.59366173,207.122779)
\lineto(340.75892787,200.19612955)
\lineto(364.30954205,189.45982971)
\lineto(383.01147591,176.6455305)
\lineto(395.1331011,158.63625176)
\lineto(399.2891074,146.86094845)
\lineto(399.2891074,130.23699885)
\lineto(396.17213102,116.38371129)
\lineto(387.51380031,99.41342089)
\lineto(377.12384126,85.90646278)
\lineto(364.65586016,75.51649239)
\lineto(351.14888693,67.89718026)
\lineto(333.13962709,63.04853018)
\lineto(307.16471055,63.74119286)
\lineto(284.99944441,68.58984294)
\lineto(255.21484724,75.17016294)
\lineto(228.54725669,75.51648105)
\lineto(207.07465701,69.97516074)
\lineto(185.60204598,61.66318971)
\lineto(165.51477921,49.88789018)
\lineto(144.73483843,38.45892011)
\lineto(127.07188913,29.45427885)
\lineto(112.52592756,24.95196199)
\lineto(98.67262866,23.22029963)
\lineto(85.16567055,24.95196199)
\lineto(74.08303748,29.1079494)
\lineto(59.53707969,37.76624987)
\lineto(46.72278047,48.8488905)
\lineto(33.21581858,69.28249806)
\lineto(24.55751055,90.40876829)
\lineto(21.09418961,113.95937113)
\lineto(25.25018079,139.24162877)
\lineto(36.33281008,159.67524011)
\lineto(49.83976819,173.87486845)
\lineto(69.92705008,187.38182656)
\lineto(93.13131969,199.15712987)
\lineto(115.29658961,207.46909711)
\closepath
}
}
{
\newrgbcolor{curcolor}{0.60000002 0.60000002 0.60000002}
\pscustom[linewidth=22.67716535,linecolor=curcolor]
{
\newpath
\moveto(211.00118736,207.21141648)
\curveto(225.11894736,207.6197642)(239.26025004,207.21279223)(253.33115902,205.99320294)
\curveto(267.3742998,204.77602121)(281.35200563,202.7485579)(295.13027335,199.77361239)
\curveto(308.36194579,196.9166864)(321.39413477,193.19004577)(334.27442453,189.02613648)
\curveto(342.63587335,186.32306751)(350.97313319,183.41952861)(358.76557414,179.35772325)
\curveto(366.55797729,175.29591412)(373.83009067,170.01854845)(379.3242387,163.16037648)
\curveto(385.03778831,156.02832105)(388.71647807,147.16109963)(389.06816311,138.02943223)
\curveto(389.35124973,130.68173207)(387.50211587,123.35027255)(384.42229225,116.67318152)
\curveto(381.34250642,109.9960905)(377.05977256,103.93382278)(372.34968752,98.28725018)
\curveto(365.35578516,89.90281806)(357.14767177,82.20251444)(347.1290513,77.86196137)
\curveto(339.87901036,74.72089806)(331.91002768,73.45256798)(324.00884973,73.43698499)
\curveto(316.10763398,73.42148892)(308.24871114,74.62803129)(300.52078673,76.27369806)
\curveto(288.53896626,78.82522688)(276.78096941,82.4381642)(264.69008311,84.40900262)
\curveto(250.68398169,86.69202625)(236.32276343,86.73730121)(222.28841004,84.63494278)
\curveto(206.67072941,82.29540011)(191.46779902,77.29881176)(177.43186201,70.06155696)
\curveto(164.20684154,63.24243129)(152.07640248,54.49094656)(139.11974358,47.17475255)
\curveto(131.69629036,42.9829768)(123.9602513,39.24800987)(115.76747146,36.89068074)
\curveto(107.57469162,34.53335538)(98.88292343,33.58645144)(90.47537193,34.99773081)
\curveto(82.18489886,36.38935664)(74.36331776,40.04839664)(67.57200941,45.00285837)
\curveto(60.78070484,49.95732011)(54.98487492,56.1829957)(49.97274894,62.93185617)
\curveto(41.9967628,73.6715727)(35.87035272,86.00033869)(33.69093351,99.19912546)
\curveto(30.78120752,116.82071538)(35.27235587,135.51008987)(45.71185697,150.00159223)
\curveto(57.89635839,166.91540514)(77.00756973,177.41095664)(96.21648185,185.50767223)
\curveto(110.48085729,191.52022404)(125.18807618,196.53185869)(140.26902232,200.02302341)
\curveto(152.66660217,202.89300388)(165.2903187,204.72992231)(177.96543681,205.86046215)
\curveto(188.94720941,206.83996451)(199.97610516,207.2909831)(211.00118736,207.21143538)
\closepath
}
}
{
\newrgbcolor{curcolor}{0.74901962 0.74901962 0.74901962}
\pscustom[linewidth=22.67716535,linecolor=curcolor]
{
\newpath
\moveto(212.68095122,231.41040559)
\curveto(228.2602961,231.24055362)(243.81731909,230.19932401)(259.3468498,228.94293346)
\curveto(272.58531405,227.8718985)(285.84162901,226.639867)(298.86300854,224.0231111)
\curveto(313.81164476,221.01906322)(328.34109358,216.21129251)(342.61738964,210.85661661)
\curveto(358.81591563,204.78096559)(375.05306838,197.79663677)(388.05169516,186.37990102)
\curveto(397.82460854,177.79630669)(405.56005421,166.72079047)(409.34007311,154.27492716)
\curveto(413.07806366,141.96750795)(412.85503374,128.60135614)(409.46933295,116.19236968)
\curveto(406.08378334,103.78337944)(399.61523531,92.31797094)(391.34929516,82.46307582)
\curveto(380.29897705,69.28859141)(365.70935815,58.76866574)(349.13068728,54.20526417)
\curveto(336.80889831,50.81359929)(323.77224949,50.76937503)(311.07957547,52.26141913)
\curveto(289.17521012,54.83631677)(267.87650271,61.88629598)(245.85154775,63.04106637)
\curveto(230.78533421,63.83098763)(215.49843405,61.79086291)(201.42517043,56.35414102)
\curveto(188.78534177,51.47117661)(177.35417957,43.97669472)(165.94207752,36.67104181)
\curveto(156.35582744,30.53423803)(146.67188791,24.47224999)(136.30454555,19.77387015)
\curveto(125.93720697,15.07549409)(114.80946523,11.75440433)(103.43348413,11.37545755)
\curveto(90.29128823,10.93767866)(77.18037925,14.47191866)(65.48989232,20.49201062)
\curveto(46.19677232,30.42713377)(30.63981358,47.16097692)(21.36604728,66.78056874)
\curveto(12.69279878,85.12970637)(9.42916161,106.04834826)(12.44030744,126.11944999)
\curveto(15.09145327,143.79091834)(22.63686807,160.79570826)(34.47781799,174.1785807)
\curveto(49.46955594,191.12253535)(70.42709673,201.56804212)(91.56444098,209.63416999)
\curveto(109.47984004,216.47078354)(127.94191752,221.90443645)(146.7560202,225.60988149)
\curveto(158.85475279,227.99272999)(171.09540287,229.66062417)(183.39339216,230.56438858)
\curveto(193.13655311,231.28040118)(202.91194209,231.51689755)(212.68080004,231.41040559)
\closepath
}
}
{
\newrgbcolor{curcolor}{0 0 0}
\pscustom[linewidth=2.66666825,linecolor=curcolor]
{
\newpath
\moveto(49.76086677,172.04476074)
\curveto(17.76758551,143.47932955)(16.39644472,105.54444199)(31.0222526,74.00795444)
\curveto(45.64806047,42.47147066)(76.26959244,17.33439601)(111.23406614,23.96196011)
\curveto(146.19854362,30.5895242)(185.5030148,68.97993176)(226.40951811,74.69312987)
\curveto(267.31602142,80.40632798)(309.82054677,53.44109585)(347.07017197,65.78187318)
\curveto(384.31979717,78.1226505)(416.31243969,129.76792168)(392.08816252,165.18924577)
\curveto(367.86388535,200.61056987)(287.42684598,219.80577365)(215.89758236,219.80577365)
\curveto(144.36831874,219.80577365)(81.75414803,200.61018814)(49.76086677,172.04476074)
\closepath
}
}
{
\newrgbcolor{curcolor}{0 0 0}
\pscustom[linewidth=0.99999871,linecolor=curcolor]
{
\newpath
\moveto(376.82121146,63.81538574)
\curveto(379.64149493,59.29215299)(383.3657658,55.3351048)(387.70991698,52.24612716)
\curveto(394.60506028,47.3432333)(403.06874532,44.68321314)(411.52899099,44.76012653)
}
}
{
\newrgbcolor{curcolor}{0 0 0}
\pscustom[linestyle=none,fillstyle=solid,fillcolor=curcolor]
{
\newpath
\moveto(375.18595489,61.14594561)
\lineto(373.48795014,69.16134369)
\lineto(379.93791751,64.10884457)
\curveto(377.81318366,64.4338909)(375.83517922,63.20058421)(375.18595489,61.14594561)
\closepath
}
}
{
\newrgbcolor{curcolor}{0 0 0}
\pscustom[linewidth=0.99999871,linecolor=curcolor]
{
\newpath
\moveto(228.05583118,242.06134782)
\lineto(226.42647685,196.1900042)
}
}
{
\newrgbcolor{curcolor}{0 0 0}
\pscustom[linewidth=0.99999871,linecolor=curcolor]
{
\newpath
\moveto(224.05835731,242.20333846)
\lineto(232.05330505,241.91935718)
}
}
{
\newrgbcolor{curcolor}{0 0 0}
\pscustom[linewidth=0.99999871,linecolor=curcolor]
{
\newpath
\moveto(222.42900298,196.33199484)
\lineto(230.42395072,196.04801356)
}
}
{
\newrgbcolor{curcolor}{0 0 0}
\pscustom[linestyle=none,fillstyle=solid,fillcolor=curcolor]
{
\newpath
\moveto(143.49462028,22.66279303)
\curveto(143.49462028,22.19996977)(143.1194279,21.82477738)(142.65660464,21.82477738)
\curveto(142.19378138,21.82477738)(141.818589,22.19996977)(141.818589,22.66279303)
\curveto(141.818589,23.12561629)(142.19378138,23.50080867)(142.65660464,23.50080867)
\curveto(143.1194279,23.50080867)(143.49462028,23.12561629)(143.49462028,22.66279303)
\closepath
}
}
{
\newrgbcolor{curcolor}{0 0 0}
\pscustom[linewidth=2.46803149,linecolor=curcolor]
{
\newpath
\moveto(143.49462028,22.66279303)
\curveto(143.49462028,22.19996977)(143.1194279,21.82477738)(142.65660464,21.82477738)
\curveto(142.19378138,21.82477738)(141.818589,22.19996977)(141.818589,22.66279303)
\curveto(141.818589,23.12561629)(142.19378138,23.50080867)(142.65660464,23.50080867)
\curveto(143.1194279,23.50080867)(143.49462028,23.12561629)(143.49462028,22.66279303)
\closepath
}
}
{
\newrgbcolor{curcolor}{0 0 0}
\pscustom[linestyle=none,fillstyle=solid,fillcolor=curcolor]
{
\newpath
\moveto(139.62687049,45.92260863)
\curveto(139.62687049,45.45978537)(139.25167811,45.08459299)(138.78885485,45.08459299)
\curveto(138.32603159,45.08459299)(137.9508392,45.45978537)(137.9508392,45.92260863)
\curveto(137.9508392,46.38543189)(138.32603159,46.76062428)(138.78885485,46.76062428)
\curveto(139.25167811,46.76062428)(139.62687049,46.38543189)(139.62687049,45.92260863)
\closepath
}
}
{
\newrgbcolor{curcolor}{0 0 0}
\pscustom[linewidth=2.46803149,linecolor=curcolor]
{
\newpath
\moveto(139.62687049,45.92260863)
\curveto(139.62687049,45.45978537)(139.25167811,45.08459299)(138.78885485,45.08459299)
\curveto(138.32603159,45.08459299)(137.9508392,45.45978537)(137.9508392,45.92260863)
\curveto(137.9508392,46.38543189)(138.32603159,46.76062428)(138.78885485,46.76062428)
\curveto(139.25167811,46.76062428)(139.62687049,46.38543189)(139.62687049,45.92260863)
\closepath
}
}
{
\newrgbcolor{curcolor}{0 0 0}
\pscustom[linewidth=0.99999871,linecolor=curcolor]
{
\newpath
\moveto(246.38887181,206.48704357)
\curveto(254.85093165,204.82848861)(262.82993008,200.77442751)(269.15897575,194.91778499)
\curveto(274.42245921,190.04717585)(278.54248441,183.94871538)(281.09665134,177.24773018)
}
}
{
\newrgbcolor{curcolor}{0 0 0}
\pscustom[linestyle=none,fillstyle=solid,fillcolor=curcolor]
{
\newpath
\moveto(247.22418037,203.47005264)
\lineto(240.20651042,207.69878)
\lineto(248.30127942,208.96548498)
\curveto(246.61371624,207.63421507)(246.16537376,205.34674136)(247.22418037,203.47005264)
\closepath
}
\rput[bl](180,135){$\Omega$}
\rput[bl](415,35){$\partial \Omega$}
\rput[bl](143,47){$x$}
\rput[bl](148,20){$y$}
\rput[bl](271,157){$2 \delta$}
}
\end{pspicture}
\caption{The area as a surface layer integral.}
\label{figarea}
\end{figure}

In order to give a first idea of our constructions, we now outline a few basic concepts.
We always assume that the measure~$\mu$ describing the whole space~$N:= \supp \mu$
is a minimizer of the causal variational principle of infinite total volume
(for details see the preliminaries in Section~\ref{secprelim}). The
corresponding Euler-Lagrange equations imply that integrating the Lagrangian over one of its arguments
gives a constant denoted by~$\s>0$, i.e.
\[ 
\ell(\x) := \int_\G \L(\x,\y)\: d\mu(\y) - \s = 0 \qquad \text{for all~$\x \in N$}\:. \]
Given a subset~$\Omega \subset N$, its {\em{volume}}~$V$ is given by~$V=\mu(\Omega)$.
In order to define the {\em{area}}~$A$ of its boundary, we consider a double integral over the Lagrangian,
where one integral is over~$\Omega$, and the other integral is over its complement~$N \setminus \Omega$,
\beq \label{Adefintro}
A(\Omega) := \int_\Omega d\mu(\x) \int_{N \setminus \Omega} d\mu(\y)\: \L(\x,\y) \:.
\eeq
The idea behind this so-called {\em{surface layer integral}} is that, in the typical situation when
the Lagrangian~$\L(x,y)$ vanishes unless~$x$ and~$y$ are close together, we only get contributions
to the double integral if both~$x$ and~$y$ are close to the boundary of~$\Omega$.
In this way, the surface layer integral can be understood as a ``smeared'' surface integral with
the width of the surface layer given by the range of the Lagrangian.
In order to make this idea precise, one could introduce a distance
function~$d \in C^0(N \times N, \R^+_0)$ on~$N$ and assume that~$\L$ vanishes
on distances larger than~$\delta$; that is,
\beq \label{shortrange}
d(x,y) > \delta \quad \Longrightarrow \quad \L(x,y) = 0 \:.
\eeq
Under this assumption,
the surface layer integral~\eqref{Adefintro} only involves pairs~$(x,y)$ of distance at most~$\delta$,
with~$x$ lying in~$\Omega$ and~$y$ lying in the complement~$M \setminus \Omega$.
This situation is illustrated in Figure~\ref{figarea}.
In the main part of the paper, we will not assume~\eqref{shortrange} because
a distance function will not be needed for the main constructions.
Instead, we shall work with the more
general assumption that the Lagrangian has compact range (see~(vi) on page~\pageref{compactrange}).
In the concrete example in Section~\ref{seciso}, however, we shall work with~\eqref{shortrange},
choosing~$d$ as the Euclidean norm in~$\R^3=\G=N$  and Lagrangian as a bump function
on the scale~$\delta$. Similarly, in Section~\ref{secriemann} we choose the
Lagrangian again as a bump function of the Riemannian distance of~$x$ and~$y$.
But our abstract results also apply to much more general situations.


Instead of working with subsets~$\Omega \subset N$, it is preferable to work more generally
with measures~$\rho$ with
\beq \label{rhoineq}
0 \leq \rho \leq \mu \:.
\eeq
In order to describe a subset~$\Omega \subset N$, we choose~$\rho$
as the measure~$\mu$ multiplied by the characteristic function of~$\Omega$,
\beq \label{nudef}
\rho := \chi_\Omega\, \mu \:.
\eeq
One advantage of working
with the measure~$\rho$ is that we can describe more general geometric situations including
boundary layers which interpolate between the interior and exterior regions.
The notions of volume and area immediately generalize to this situation:
\begin{Def} \label{defarea} {\bf{(Area and Volume)}}
Given a measure~$\rho$ satisfying~\eqref{rhoineq}, we define its
area~$A$ and volume~$V$ by
\begin{align*}
A(\rho) &:= \int_N d\rho(\x) \int_N d(\mu-\rho)(\y)\: \L(\x,\y) \\
V(\rho) &:= \rho(N) \:.
\end{align*}
\end{Def} \noindent
Moreover, a notion of {\em{mean curvature}} can be introduced
via the first variation of the area functional by
\beq \label{Hdef}
H(x) = \s - 2 \int_N \L(x,y)\: d\rho(y)
\eeq
(for motivation and details see again Section~\ref{seciso}).
A further advantage of working with the measure~$\rho$ is that
the powerful tools of measure theory become available to our problem.
More specifically, in Section~\ref{secminmov} we set up a minimizing movement scheme where, in each step, the area is minimized for fixed volume, subject to certain constraints (for details see~\eqref{Shdef}).
For the penalization, we prefer not to work with the Wasserstein distance, but rather with
a double integral involving the total variation measure (see the last summand~\eqref{Shdef}).
This is sufficient for our purposes and is easier because we do not need to introduce methods and results from optimal transport.
Our constructions give a Lipschitz curve of measures~$\rho(t)$ which obeys the so-called
{\em{isoperimetric flow}} (Theorem~\ref{thmlip}).

Clearly, it is an important task analyze how the isoperimetric flow relates to the classical
mean curvature flow. This connection is worked out in Sections~\ref{seceuclid} and~\ref{secriemann}.
To this end, as in Section~\ref{seciso}, we choose the Lagrangian as a bump function on a scale~$\delta$
and analyze the asymptotic behavior as $\delta \searrow 0$.
In this limiting case, we recover the classical mean curvature functional, up to higher order corrections
in~$\delta$ which involve derivatives of curvature and higher order curvature terms.
These results establish that the isoperimetric flow can indeed be regarded as a regularized version
of the classical mean curvature flow.

As an application, in Section~\ref{secscalar} we use the isoperimetric flow to introduce notions
of {\em{local dimension}} and {\em{scalar curvature}}. For this purpose, we consider the isoperimetric flow emanating from a single point. We thus obtain measures describing small 
isoperimetric balls. Comparing their area and volume with that
of small geodesic balls in a Riemannian manifold gives us information on 
the local dimension and scalar curvature.

\subsection{Connections with Metric Measure Spaces  and Nonlocal Perimeter}
The present construction belongs to a broader effort to develop geometry and
analysis beyond smooth manifolds. In \emph{metric measure geometry}, one starts
from a triple $(X,d,m)$ consisting of a metric space and a reference measure.
This setting supports notions of Sobolev regularity, perimeter, curvature and
geometric evolution under suitable structural hypotheses; see, for example,
\cite{ambrosio+tilli,shioya,sturm-mms}. Our starting point is different:
we are given the measure~$\mu$ and the interaction kernel~$\L$, but do not
require a prescribed metric on~$N=\supp\mu$. Thus the interaction kernel,
rather than a distance function, is the primary source of geometric information.
This distinction is especially relevant for spaces with singular or discrete
microstructure, where a classical infinitesimal metric description may be
unavailable or inappropriate.

\emph{Nonlocal perimeter and BV theory.}
If $\L$ is symmetric, the area in~\eqref{Adefintro} can be rewritten as
\[
 A(\Omega)=\frac12\int_N\!\int_N
 |\chi_\Omega(x)-\chi_\Omega(y)|\,\L(x,y)\,
 d\mu(x)\,d\mu(y).
\]
This is a nonlocal perimeter functional: it measures the interaction of pairs
of points separated by the boundary, without assigning a surface measure to
that boundary. The formula places our construction alongside nonlocal
approximations of total variation and perimeter, including the foundational
work of Bourgain, Brezis and Mironescu~\cite{BBM}, the characterization of
Sobolev and BV regularity by nonlocal difference quotients due to
Ponce~\cite{ponce}, and the theory of nonlocal perimeters developed by
Caffarelli, Roquejoffre and Savin~\cite{CRS}. In metric measure spaces,
perimeter can instead be defined by relaxation of the integral of upper
gradients; its structure and relation to fine properties of sets of finite
perimeter are studied, for instance, in~\cite{miranda}. The two viewpoints
are complementary. A suitable family of increasingly concentrated kernels
may approximate a local perimeter, but this requires assumptions on the
kernels and the underlying geometry. Our general construction does not rely
on such a limiting procedure: the interaction scale is retained as part of
the geometric data.

\emph{Synthetic curvature.}
A major achievement of metric measure geometry is the formulation of lower
Ricci curvature bounds without a smooth metric tensor. In the approaches of
Lott--Villani and Sturm, curvature-dimension conditions are expressed through
the displacement convexity of entropy along Wasserstein geodesics
\cite{lott-villani,sturmI,sturmII}. The $RCD(K,N)$ theory supplements these
conditions with infinitesimal Hilbertianity and provides a robust setting for
second-order calculus and geometric analysis~\cite{AGS-RCD,gigli-nonsmooth}.
Our approach addresses a different curvature quantity and uses different
data: rather than encoding Ricci curvature through optimal transport, we
investigate scalar curvature through the small-volume behavior of area and
volume along an isoperimetric evolution. On a smooth Riemannian manifold,
the classical expansion of the volume of a geodesic ball contains scalar
curvature in its first non-Euclidean correction~\cite{gray}. This motivates
our synthetic construction. It does not, by itself, identify our curvature
with a curvature-dimension invariant, nor does it imply that arbitrary
metric measure spaces admit a compatible interaction kernel. Establishing
such connections would be an interesting separate problem.

\emph{Geometric flows without a prescribed metric.}
Metric gradient-flow theory, especially the minimizing-movement framework,
provides a natural language for variational evolutions in non-smooth settings
\cite{degiorgi-minmov,ambrosio+gigli,braides-minmov}. We use this philosophy,
but the time-discrete evolution is formulated directly for measures and the
interaction-based area functional. In particular, our construction does not
require a Wasserstein geometry on the space of measures. This leads to a
Lipschitz evolution even when the support of the reference measure lacks a
smooth boundary calculus. The resulting flow is not claimed to coincide with
metric-measure mean curvature flows in general; rather, our smooth examples
show how it recovers a regularized mean curvature description, with explicit
finite-range corrections.

These comparisons highlight the central feature of our approach: the kernel
$\L$ simultaneously supplies a notion of interaction range, a nonlocal
boundary energy and, through the isoperimetric flow, a means of extracting
local geometric information. In contrast to approaches that begin with a
metric and then develop analysis on the resulting space, we study to which extent
the geometry 
is encoded in the variational interaction itself.

\subsection{Structure of the Paper}
The paper is organized as follows. Section~\ref{secprelim} provides the necessary
preliminaries on causal variational principles. In Section~\ref{seciso} we introduce the problem
and motivate the isoperimetric flow. In Section~\ref{secminmov} the isoperimetric
flow is constructed and its properties are studied; this is the technical core of the paper.
We proceed by studying the examples of the isoperimetric flow in Euclidean
space (Section~\ref{seceuclid}) and on a Riemannian manifold (Section~\ref{secriemann}).
It is shown that our functional coincides with the standard area up to
curvature corrections which we work out (see Theorems~\ref{thmeuclid} and~\ref{thmriemann}).
In Section~\ref{secscalar} it is explained how the isoperimetric flow can be used to
define notions of local dimension and scalar curvature.
Finally, in Section~\ref{secquasi} we relate our constructions and results
to the synthetic scalar curvature defined previously in the context of a positive quasi-local mass~\cite{matter}.

\section{Preliminaries on Causal Variational Principles} \label{secprelim}
We consider causal variational principles in the non-compact setting as
introduced in~\cite[Section~2]{jet}. Thus we let~$\G$ be a (possibly non-compact)
smooth manifold of dimension~$m \geq 1$. 
Moreover, we are given a non-negative function~$\L : \G \times \G \rightarrow \R^+_0$
(the {\em{Lagrangian}}) with the following properties:
\begin{itemize}[leftmargin=2em]
\item[(i)] $\L$ is symmetric: $\L(\x,\y) = \L(\y,\x)$ for all~$\x,\y \in \G$.\label{Cond1}
\item[(ii)] $\L$ is continuous, i.e.\ for all sequences~$\x_n \rightarrow \x$ and~$\y_{n'} \rightarrow \y$,
\label{Cond2}%
\[ \L(\x,\y) = \lim_{n,n' \rightarrow \infty} \L(\x_n, \y_{n'})\:. \]
\end{itemize}
We denote the (positive) Borel measures on~$\G$
by~$\meas$. The {\em{causal variational principle}} is to minimize the action
\beq \label{Sact} 
\Sact (\mu) = \int_\G d\mu(\x) \int_\G d\mu(\y)\: \L(\x,\y) 
\eeq
under variations of the measure~$\mu \in \meas$, keeping the total volume~$\mu(\G)$ fixed
({\em{volume constraint}}).

If the total volume~$\mu(\G)$ is finite, one minimizes~\eqref{Sact}
over all regular Borel measures with the same total volume.
If the total volume~$\mu(\G)$ is infinite, however, it is not obvious how to implement the volume 
constraint, making it necessary to proceed as follows.
We need to impose additional assumptions:
\begin{itemize}[leftmargin=2em]
\item[(iii)] The measure~$\mu$ is {\em{locally finite}}
(meaning that any~$\x \in \G$ has an open neighborhood~$U$ with~$\mu(U)< \infty$)
and {\em{regular}} (meaning that the measure of a set can be recovered by approximation from inside
with compact and from outside with open sets). \label{Cond3}
\item[(iv)] The function~$\L(\x,.)$ is $\mu$-integrable for all~$\x \in \G$, giving
a continuous and bounded function on~$\G$. \label{Cond4}
\end{itemize}
Given a regular Borel measure~$\mu$ on~$\G$, we vary over all
regular Borel measures~$\tilde{\mu}$ with
\beq \label{totvol}
\big| \tilde{\mu} - \mu \big|(\G) < \infty \qquad \text{and} \qquad
\big( \tilde{\mu} - \mu \big) (\G) = 0
\eeq
(where~$|.|$ denotes the total variation of a measure).
We then consider the difference of the actions defined by
\begin{align*}
\big( &\Sact(\tilde{\mu}) - \Sact(\mu) \big) = \int_\G d(\tilde{\mu} - \mu)(\x) \int_\G d\mu(\y)\: \L(\x,\y) \\
&\quad + \int_\G d\mu(\x) \int_\G d(\tilde{\mu} - \mu)(\y)\: \L(\x,\y) 
+ \int_\G d(\tilde{\mu} - \mu)(\x) \int_\G d(\tilde{\mu} - \mu)(\y)\: \L(\x,\y) \:.
\end{align*}
The measure~$\mu$ is said to be a {\em{minimizer}} of the causal action
with respect to {\em{variations of finite volume}}
if this difference is non-negative for all~$\tilde{\mu}$ satisfying~\eqref{totvol},
\[ 
\big( \Sact(\tilde{\mu}) - \Sact(\mu) \big) \geq 0 \:. \]
These variations of the causal action are well-defined.
The existence theory for minimizers is developed in~\cite{noncompact}.
It is shown in~\cite[Lemma~2.3]{jet} that a minimizer
satisfies the {\em{Euler-Lagrange (EL) equations}}
which state that for a suitable value of the parameter~$\s>0$,
the lower semi-continuous function~$\ell : \G \rightarrow \R_0^+$ defined by
\beq \label{ldef}
\ell(\x) := \int_\G \L(\x,\y)\: d\mu(\y) - \s
\eeq
is minimal and vanishes on the support of~$\mu$,
\beq \label{EL}
\ell|_N \equiv \inf_\G \ell = 0 \:.
\eeq
For further details we refer to~\cite[Section~2]{jet}.

Finally, as in~\cite{nonlocal} we make the following simplifying assumptions:
\begin{itemize}[leftmargin=2em]
\item[(v)] The Lagrangian is strictly positive on the diagonal, i.e.\
\[ 
\L(x,x) > 0 \qquad \text{for all~$\x \in \G$} \:. \]
\item[(vi)] The Lagrangian has {\em{compact range}}, meaning that for every compact set~$K \subset \G$ there is
a compact set~$K' \subset \G$ such that
\beq \label{compactrange}
\L(\x,\y) = 0 \qquad \text{for all~$\x \in K$ and~$\y \not \in K'$}\:.
\eeq
\end{itemize}

\section{Motivation of the Isoperimetric Flow} \label{seciso}
Causal variational principles are a general class of variational principles.
A main difference compared to most geometric variational principles is that,
instead of taking integrals of derivatives of fields or tensors, the action is formulated with a nonlocal kernel~$\L(x,y)$
(see~\eqref{Sact}). In simple terms, this means that all geometric quantities are ``smeared out''
on small scales, having the main advantage that the methods and results also apply to non-smooth
situations. On the other hand, the setup makes it necessary to formulate the geometric objects
purely in terms of the Lagrangian~$\L(x,y)$. It is not obvious how to recover the usual geometric
quantities like mean curvature and how to formulate corresponding curvature flows.
We now explain how this can be done. For pedagogical reasons, we consider the simplest possible
setting of a rotationally and translationally invariant Lagrangian in three-dimensional Euclidean space.
This makes it possible to explain the ideas and concepts in a simple setting. This will also
serve as a motivation for the general construction of the isoperimetric flow using minimizing 
movements in Section~\ref{secminmov}.

We consider the case that~$\G=\R^3$ with the Euclidean norm~$\| \cdot \|$.
We choose the Lagrangian as a bump function on the scale~$\delta$. More precisely, we choose
\beq \label{Leuclid}
\L(\x,\y) = \eta \Big( \frac{\|x-y\|}{\delta} \Big) \:,
\eeq
where~$\eta \in C^\infty(\R)$ is a non-negative test function with
\beq \label{etaprop}
\eta|_{\big(-\frac{1}{2},\frac{1}{2} \big)}=1 \qquad \text{and} \qquad \supp \eta \subset (-1,1) \:,
\eeq
and~$\delta>0$ is a parameter.
Moreover, we let~$\mu$ be the Lebesgue measure on~$\G$.
Then~$N:= \supp \mu = \G = \R^3$.
The parameter~$\delta$ describes the length scale of the non-locality of the Lagrangian.
We also refer to it as the {\em{range}} of the Lagrangian.
We are interested in the situation that the range~$\delta$ is very small.
This can be made precise by analyzing the asymptotics~$\delta \searrow 0$.
In this section, we anticipate this analysis by working with error terms of the form~$\O(\delta^p)$
with~$p \in \N$; the details of the asymptotic expansion will be postponed to Section~\ref{seceuclid}.
With these choices, all the conditions in Section~\ref{secprelim} are satisfied.
Moreover, the EL equations~\eqref{ldef} are satisfied with
\beq \label{seuclid}
\s = \int_N \eta \Big( \frac{\|y\|}{\delta} \Big)\: d^3 y
= \delta^3\: \int_{\R^3} \eta \big( \|\xi\| \big)\: d^3 \xi \:.
\eeq

Next, we consider an open subset~$\Omega \subset N$, for simplicity relatively compact
and with smooth boundary~$\partial \Omega$. Let us analyze how to describe the geometry of~$\Omega$
in our setting. The first quantity of interest is the {\em{volume}}~$V$. It is given in an obvious way
by~$V = \mu(\Omega)$. For the later constructions, it is preferable to describe the set~$\Omega$ by
the measure obtained by multiplying~$\mu$ with the characteristic function of~$\Omega$~\eqref{nudef}.
Then the volume of~$\Omega$ is the total volume of~$\rho$,
\[ V(\Omega) = \rho(N) \:. \]
The next quantity of interest is the {\em{area}}~$A$ of~$\partial \Omega$.
This cannot be introduced in an obvious way, because we have no measure on the boundary.
Instead, we work with the concept of a {\em{surface layer integral}}, being a double integral of the Lagrangian,
where one variable is inside~$\Omega$ and the other outside,
\beq \label{Adef}
A(\Omega) := \int_\Omega d\mu(\x) \int_{N \setminus \Omega} d\mu(\y)\: \L(\x,\y) \:.
\eeq
If~$\delta$ is small, the integrand vanishes unless both~$x$ and~$y$ are close to the boundary
(see Figure~\ref{figarea}; for a more detailed introduction to the concept of a surface layer integral
see~\cite[Section~9.1]{intro}). 
Asymptotically for small~$\delta$, we have the expansion
(for the proof see Theorem~\ref{thmeuclid})
\beq \label{Aasy}
A(\Omega) = \delta^4\, \s_1\, \mu_2(\partial \Omega) + \O\big( \delta^6 \big) 
\eeq
with
\beq \label{s1def}
\s_1 := 2 \int_{\R^3} \big|\xi_3 \big| \: \eta \big( \|\xi\| \big) \: d^3 \xi \:,
\eeq
where~$\xi_3$ is the third component of~$\xi \in \R^3$, and $\mu_2$ denotes the two-dimensional
Hausdorff measure. This formula can be derived by rewriting the surface layer integrals 
using Fubini as a surface integral over~$\partial \Omega$ times four transverse integrals.
For small~$\delta$, the transverse integrals can be carried out to obtain a constant
(this method will be developed systematically in Section~\ref{seceuclid}).

\begin{figure}
\psset{xunit=.4pt,yunit=.4pt,runit=.4pt}
\begin{pspicture}(389.97407688,225.24889368)
{
\newrgbcolor{curcolor}{0.89803922 0.89803922 0.89803922}
\pscustom[linestyle=none,fillstyle=solid,fillcolor=curcolor]
{
\newpath
\moveto(253.57548472,224.4986688)
\lineto(314.34333354,224.4986688)
\lineto(330.29829165,215.63478092)
\lineto(344.03732787,204.11173084)
\lineto(355.11720189,192.1454858)
\lineto(366.64022551,175.30410376)
\lineto(375.06093732,158.90591667)
\lineto(381.26563654,140.73494596)
\lineto(383.48161134,123.00717777)
\lineto(383.48161134,107.05217809)
\lineto(381.70886173,86.66523635)
\lineto(375.94731213,66.27830974)
\lineto(371.07217512,52.5392773)
\lineto(362.2083137,37.91387006)
\lineto(350.24206488,23.28845903)
\lineto(338.71900346,12.20860013)
\lineto(325.42316976,1.57193998)
\lineto(254.17435087,2.09212769)
\closepath
}
}
{
\newrgbcolor{curcolor}{0.89803922 0.89803922 0.89803922}
\pscustom[linewidth=0.99999871,linecolor=curcolor]
{
\newpath
\moveto(253.57548472,224.4986688)
\lineto(314.34333354,224.4986688)
\lineto(330.29829165,215.63478092)
\lineto(344.03732787,204.11173084)
\lineto(355.11720189,192.1454858)
\lineto(366.64022551,175.30410376)
\lineto(375.06093732,158.90591667)
\lineto(381.26563654,140.73494596)
\lineto(383.48161134,123.00717777)
\lineto(383.48161134,107.05217809)
\lineto(381.70886173,86.66523635)
\lineto(375.94731213,66.27830974)
\lineto(371.07217512,52.5392773)
\lineto(362.2083137,37.91387006)
\lineto(350.24206488,23.28845903)
\lineto(338.71900346,12.20860013)
\lineto(325.42316976,1.57193998)
\lineto(254.17435087,2.09212769)
\closepath
}
}
{
\newrgbcolor{curcolor}{0.89803922 0.89803922 0.89803922}
\pscustom[linestyle=none,fillstyle=solid,fillcolor=curcolor]
{
\newpath
\moveto(1.34640732,221.88347414)
\lineto(121.00989695,221.88347414)
\lineto(121.00989695,5.25994379)
\lineto(1.34640732,5.25994379)
\closepath
}
}
{
\newrgbcolor{curcolor}{0.89803922 0.89803922 0.89803922}
\pscustom[linewidth=2.6928189,linecolor=curcolor]
{
\newpath
\moveto(1.34640732,221.88347414)
\lineto(121.00989695,221.88347414)
\lineto(121.00989695,5.25994379)
\lineto(1.34640732,5.25994379)
\closepath
}
}
{
\newrgbcolor{curcolor}{0 0 0}
\pscustom[linewidth=2.66666825,linecolor=curcolor]
{
\newpath
\moveto(123.03716031,223.39342155)
\lineto(122.90627528,2.43646155)
}
}
{
\newrgbcolor{curcolor}{0 0 0}
\pscustom[linewidth=2.66666825,linecolor=curcolor]
{
\newpath
\moveto(316.16521701,224.07168729)
\curveto(353.7861052,204.06386619)(380.22484535,164.40575217)(384.22400126,121.98346556)
\curveto(386.43370205,98.54337611)(382.08773669,74.51979186)(371.80602331,53.3394517)
\curveto(361.52430992,32.15910776)(345.33685795,13.88382965)(325.55261858,1.12040666)
}
}
{
\newrgbcolor{curcolor}{0 0 0}
\pscustom[linestyle=none,fillstyle=solid,fillcolor=curcolor]
{
\newpath
\moveto(128.00739004,106.82084823)
\curveto(128.00739004,104.32674229)(125.98551789,102.30486993)(123.49141219,102.30486993)
\curveto(120.9973065,102.30486993)(118.97543435,104.32674229)(118.97543435,106.82084823)
\curveto(118.97543435,109.31495417)(120.9973065,111.33682652)(123.49141219,111.33682652)
\curveto(125.98551789,111.33682652)(128.00739004,109.31495417)(128.00739004,106.82084823)
\closepath
}
}
{
\newrgbcolor{curcolor}{0 0 0}
\pscustom[linewidth=2.30662676,linecolor=curcolor]
{
\newpath
\moveto(128.00739004,106.82084823)
\curveto(128.00739004,104.32674229)(125.98551789,102.30486993)(123.49141219,102.30486993)
\curveto(120.9973065,102.30486993)(118.97543435,104.32674229)(118.97543435,106.82084823)
\curveto(118.97543435,109.31495417)(120.9973065,111.33682652)(123.49141219,111.33682652)
\curveto(125.98551789,111.33682652)(128.00739004,109.31495417)(128.00739004,106.82084823)
\closepath
}
}
{
\newrgbcolor{curcolor}{0 0 0}
\pscustom[linestyle=none,fillstyle=solid,fillcolor=curcolor]
{
\newpath
\moveto(388.82075799,107.64204021)
\curveto(388.82075799,105.14793427)(386.79888583,103.12606192)(384.30478014,103.12606192)
\curveto(381.81067445,103.12606192)(379.7888023,105.14793427)(379.7888023,107.64204021)
\curveto(379.7888023,110.13614615)(381.81067445,112.15801851)(384.30478014,112.15801851)
\curveto(386.79888583,112.15801851)(388.82075799,110.13614615)(388.82075799,107.64204021)
\closepath
}
}
{
\newrgbcolor{curcolor}{0 0 0}
\pscustom[linewidth=2.30662676,linecolor=curcolor]
{
\newpath
\moveto(388.82075799,107.64204021)
\curveto(388.82075799,105.14793427)(386.79888583,103.12606192)(384.30478014,103.12606192)
\curveto(381.81067445,103.12606192)(379.7888023,105.14793427)(379.7888023,107.64204021)
\curveto(379.7888023,110.13614615)(381.81067445,112.15801851)(384.30478014,112.15801851)
\curveto(386.79888583,112.15801851)(388.82075799,110.13614615)(388.82075799,107.64204021)
\closepath
}
\rput[bl](10,200){$\Omega$}
\rput[bl](263,200){$\Omega$}
\rput[bl](100,102){$x$}
\rput[bl](360,102){$x$}
}
\end{pspicture}
\caption{The mean curvature functional.}
\label{figmean}
\end{figure}
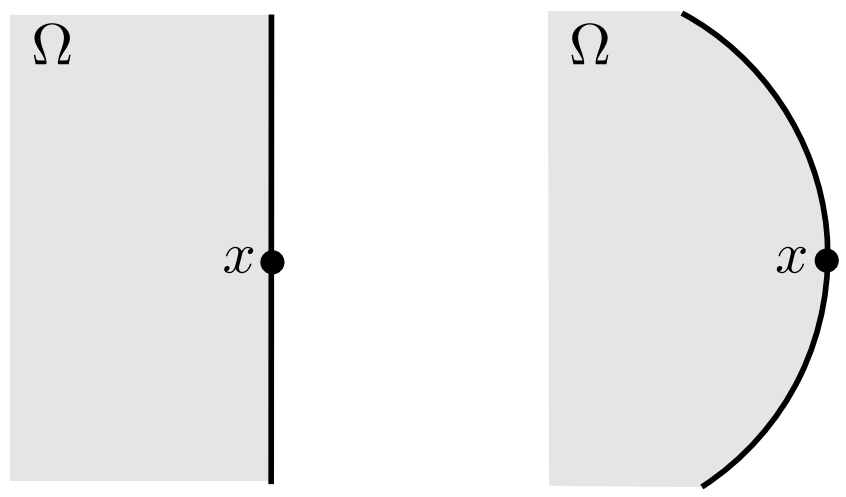

Finally, we need to introduce the {\em{mean curvature}}~$H$ of~$\partial \Omega$.
To this end, we choose~$x \in \partial \Omega$ and consider the integral
\beq \label{Hint}
\int_\Omega \L(x,y)\: d^3y \:.
\eeq
If~$\partial \Omega$ is flat, we may replace~$\Omega$ by the half plane 
(see the left of Figure~\ref{figmean}).
Comparing with~\eqref{seuclid}, we obtain by symmetry
\[ \int_\Omega \L(x,y)\: d^3y = \int_{\xi^3 < 0} \eta \Big( \frac{\|\xi\|}{\delta} \Big)\: d^3\xi = \frac{\s}{2} \:. \]
If~$\partial \Omega$ is curved, however, the integral in~\eqref{Hint} changes. It becomes smaller if
the mean curvature is positive (see the right of Figure~\ref{figmean}). This leads us to define
mean curvature by~\eqref{Hdef}.
In Section~\ref{seceuclid} it is shown that this definition indeed agrees with the differential geometric
definition up to a constant, i.e.\
\[ H(x) = \delta^4\,\s_1 \Tr(W_x) + \O\big( \delta^6 \big) \:, \]
where~$W_x : T_x \partial \Omega \rightarrow T_x \partial \Omega$ is the Weingarten map.

\begin{figure}
\psset{xunit=.5pt,yunit=.5pt,runit=.5pt}
\begin{pspicture}(498.71153343,281.80055633)
{
\newrgbcolor{curcolor}{0.89803922 0.89803922 0.89803922}
\pscustom[linestyle=none,fillstyle=solid,fillcolor=curcolor]
{
\newpath
\moveto(55.07676472,195.82757114)
\curveto(23.08348346,167.26213996)(21.71234268,129.3272524)(36.33815055,97.79076484)
\curveto(50.96395843,66.25428106)(81.58549039,41.11720641)(116.54996409,47.74477051)
\curveto(151.51444157,54.3723346)(190.81891276,92.76274216)(231.72541606,98.47594027)
\curveto(272.63191937,104.18913838)(315.13644472,77.22390625)(352.38606992,89.56468358)
\curveto(389.63569512,101.9054609)(421.62833764,153.55073208)(397.40406047,188.97205618)
\curveto(373.17978331,224.39338027)(292.74274394,243.58858405)(221.21348031,243.58858405)
\curveto(149.68421669,243.58858405)(87.07004598,224.39299854)(55.07676472,195.82757114)
\closepath
}
}
{
\newrgbcolor{curcolor}{0 0 0}
\pscustom[linewidth=3.33333166,linecolor=curcolor]
{
\newpath
\moveto(55.07676472,195.82757114)
\curveto(23.08348346,167.26213996)(21.71234268,129.3272524)(36.33815055,97.79076484)
\curveto(50.96395843,66.25428106)(81.58549039,41.11720641)(116.54996409,47.74477051)
\curveto(151.51444157,54.3723346)(190.81891276,92.76274216)(231.72541606,98.47594027)
\curveto(272.63191937,104.18913838)(315.13644472,77.22390625)(352.38606992,89.56468358)
\curveto(389.63569512,101.9054609)(421.62833764,153.55073208)(397.40406047,188.97205618)
\curveto(373.17978331,224.39338027)(292.74274394,243.58858405)(221.21348031,243.58858405)
\curveto(149.68421669,243.58858405)(87.07004598,224.39299854)(55.07676472,195.82757114)
\closepath
}
}
{
\newrgbcolor{curcolor}{0 0 0}
\pscustom[linewidth=2.66666825,linecolor=curcolor]
{
\newpath
\moveto(46.1305629,227.95545309)
\curveto(8.45356762,196.48370884)(-5.20058041,155.02667955)(4.17623471,108.6222071)
\curveto(13.55304983,62.21773088)(45.95943723,10.87056994)(89.22434683,2.60382096)
\curveto(132.48925644,-5.66292423)(186.60753676,29.15173277)(233.21854904,33.2472364)
\curveto(279.82955754,37.34274002)(318.92774967,10.71859892)(360.00100951,39.26252081)
\curveto(401.07426935,67.8064427)(444.11772132,151.51506002)(418.9802725,204.56703356)
\curveto(393.84286148,257.61901088)(300.52757329,280.00804773)(223.01672353,280.46021529)
\curveto(145.50587754,280.91238285)(83.80755817,259.42719734)(46.1305629,227.95545309)
\closepath
}
}
{
\newrgbcolor{curcolor}{0 0 0}
\pscustom[linewidth=2.00000125,linecolor=curcolor]
{
\newpath
\moveto(405.12268724,165.44353334)
\lineto(486.17624315,173.68626342)
}
}
{
\newrgbcolor{curcolor}{0 0 0}
\pscustom[linestyle=none,fillstyle=solid,fillcolor=curcolor]
{
\newpath
\moveto(482.82403923,178.9742477)
\lineto(498.71159838,174.96104462)
\lineto(483.95717808,167.83170972)
\curveto(486.41278588,171.36025862)(485.94111684,175.99834005)(482.82403923,178.9742477)
\closepath
}
}
{
\newrgbcolor{curcolor}{0 0 0}
\pscustom[linewidth=1.33333413,linecolor=curcolor]
{
\newpath
\moveto(403.46634563,179.53187707)
\curveto(407.20896082,180.53470776)(411.20453766,179.5569963)(414.06122217,176.9393272)
\curveto(416.91790667,174.3216581)(418.24013506,170.42649879)(417.56731081,166.61072293)
}
}
{
\newrgbcolor{curcolor}{0 0 0}
\pscustom[linestyle=none,fillstyle=solid,fillcolor=curcolor]
{
\newpath
\moveto(236.67510663,98.28319416)
\curveto(236.67510663,95.15213102)(234.13687844,92.61390283)(231.00581529,92.61390283)
\curveto(227.87475214,92.61390283)(225.33652395,95.15213102)(225.33652395,98.28319416)
\curveto(225.33652395,101.41425731)(227.87475214,103.9524855)(231.00581529,103.9524855)
\curveto(234.13687844,103.9524855)(236.67510663,101.41425731)(236.67510663,98.28319416)
\closepath
}
}
{
\newrgbcolor{curcolor}{0 0 0}
\pscustom[linewidth=1.33333041,linecolor=curcolor]
{
\newpath
\moveto(236.67510663,98.28319416)
\curveto(236.67510663,95.15213102)(234.13687844,92.61390283)(231.00581529,92.61390283)
\curveto(227.87475214,92.61390283)(225.33652395,95.15213102)(225.33652395,98.28319416)
\curveto(225.33652395,101.41425731)(227.87475214,103.9524855)(231.00581529,103.9524855)
\curveto(234.13687844,103.9524855)(236.67510663,101.41425731)(236.67510663,98.28319416)
\closepath
}
}
{
\newrgbcolor{curcolor}{0 0 0}
\pscustom[linestyle=none,fillstyle=solid,fillcolor=curcolor]
{
\newpath
\moveto(411.38208982,171.72914777)
\curveto(411.38208982,171.20730392)(410.95905179,170.78426588)(410.43720793,170.78426588)
\curveto(409.91536407,170.78426588)(409.49232604,171.20730392)(409.49232604,171.72914777)
\curveto(409.49232604,172.25099163)(409.91536407,172.67402966)(410.43720793,172.67402966)
\curveto(410.95905179,172.67402966)(411.38208982,172.25099163)(411.38208982,171.72914777)
\closepath
}
}
{
\newrgbcolor{curcolor}{0 0 0}
\pscustom[linewidth=1.33333041,linecolor=curcolor]
{
\newpath
\moveto(411.38208982,171.72914777)
\curveto(411.38208982,171.20730392)(410.95905179,170.78426588)(410.43720793,170.78426588)
\curveto(409.91536407,170.78426588)(409.49232604,171.20730392)(409.49232604,171.72914777)
\curveto(409.49232604,172.25099163)(409.91536407,172.67402966)(410.43720793,172.67402966)
\curveto(410.95905179,172.67402966)(411.38208982,172.25099163)(411.38208982,171.72914777)
\closepath
}
}
{
\newrgbcolor{curcolor}{0 0 0}
\pscustom[linewidth=1.33333413,linecolor=curcolor]
{
\newpath
\moveto(230.94890079,98.28521019)
\lineto(240.13028787,34.02886121)
}
\rput[bl](190,170){$\Omega$}
\rput[bl](400,60){$\partial \tilde{\Omega}_t$}
\rput[bl](450,150){$n$}
\rput[bl](245,52){$g_t(x)$}
\rput[bl](225,110){$x$}
}
\end{pspicture}
\caption{A flow of surfaces.}
\label{figgraph}
\end{figure}
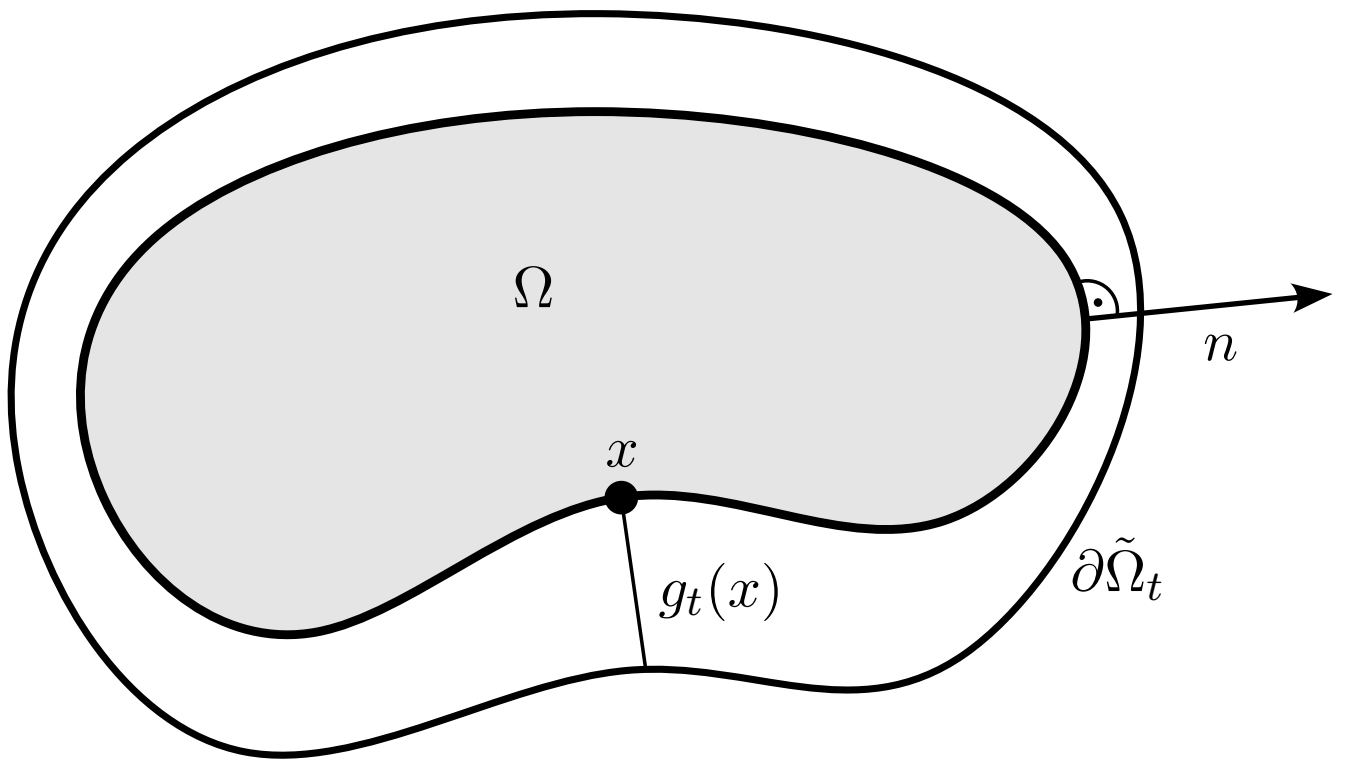

Having introduced all the needed objects, we are now in the position to formulate an analog of
the mean curvature flow. To this end, we consider a family~$(\tilde{\Omega}_t)$
of sets with~$\tilde{\Omega}_0=\Omega$. Let us begin with the simple situation
that the boundary of these sets can be parametrized
as a graph over~$\partial \Omega$, i.e.\ (see Figure~\ref{figgraph})
\[ \partial \tilde{\Omega}_t = \{ x + n \,g_t(x) \:n \:|\: x \in \partial \Omega \} \:, \]
where~$n$ is the outer normal on~$\partial \Omega$, and~$g_t$ is the height function,
\[ g_t \::\: \partial \Omega \rightarrow \R \]
(clearly, $g_0$ vanishes identically). We formulate the flow equation
\beq \label{flowsmooth}
\frac{d}{dt} g_t(x) \Big|_{t=0} = -H(x) + \alpha \:,
\eeq
where~$\alpha$ is a real parameter which determines how the volume changes.
By direct computation, one verifies that this evolution equation is the gradient flow corresponding to 
the functional
\beq \label{Ssmooth}
\Sact(\tilde{\Omega}) = -\int_{\tilde{\Omega}} d\mu(x) \int_{\tilde{\Omega}} d\mu(y)\: \L(x,y)
+ a\: \mu \big( \tilde{\Omega} \big) \qquad \text{with} \qquad a := \s - \alpha\:.
\eeq
Indeed,
\begin{align*}
\delta \Sact \big( \tilde{\Omega}_t \big) &= 
-2 \int_{\partial \Omega} d\mu_{\partial \Omega}(x)\: \delta g_t(x) \int_{\tilde{\Omega}} d\mu(y)\: \L(x,y)
+ a\: \int_{\partial \Omega} \delta g_t(x)\:d\mu_{\partial \Omega}(x) \\
&= \int_{\partial \Omega} \delta g_t(x) \:
\bigg( -2 \int_{\tilde{\Omega}} d\mu(y)\: \L(x,y)
+ \s - \alpha \bigg)\:d\mu_{\partial \Omega}(x) \\
&= \int_{\partial \Omega} \delta g_t(x) \:
\big(H(x) - \alpha \big)\:d\mu_{\partial \Omega}(x) \:.
\end{align*}

Clearly, the resulting gradient flow is well-defined only as long as the
surfaces~$\partial \tilde{\Omega}_t$ remain smooth. In order to extend the flow to the non-smooth
setting, one can use a measure-theoretic formulation. First, using~\eqref{nudef}, it is preferable to
rewrite~\eqref{Ssmooth} as
\[ \Sact(\tilde{\nu}) = -\int_N d\tilde{\nu}(x) \int_N d\tilde{\nu}\: \L(x,y)
+ b\: \tilde{\nu}(N) \:, \]
making it possible to evaluate the functional for more general measures (i.e.\ measure which are
not necessarily of the form~\eqref{nudef}). Moreover, in order to avoid differentiability issues,
it is preferable to formulate the gradient flow with the method of minimizing movements.
This will be explained in the next section.

\section{The Isoperimetric Flow with Minimizing Movements} \label{secminmov}

\subsection{The Variational Principle with Penalization}
Let~$\nu$ be a regular Borel measure with the properties
\beq \label{nuprop}
\nu \leq \mu \qquad \text{and} \qquad \text{$K := \supp \nu \subset N$ is compact} \:.
\eeq
Given a parameter~$h>0$, we consider a regular Borel measure~$\rho$ on~$N$
satisfying the constraint
\beq \label{rhoprop}
\rho \leq \mu \qquad \text{and} \qquad \rho(N) < \infty\:.
\eeq
Since~$\mu$ is locally finite and~$K$ is compact, it follows from the first condition in~\eqref{nuprop}
that the measure~$\nu$ has finite total volume. In view of the second condition in~\eqref{rhoprop},
the same is true for the measure~$\rho$. Therefore, the difference~$\rho-\nu$ 
of these measures is a signed measure. We denote its total variation measure by~$|\rho-\nu|$.
Given parameter~$a > 0$ and a compactly supported continuous function~$\phi \in C^0_0(N, \R)$,
for any~$h>0$ we consider the action
\beq \label{Shdef}
\begin{split}
\Sact^h(\rho) &:= -\int_N d\rho(\x) \int_N d\nu(\y)\: \L(\x,\y) + \int_N \phi\: d\rho + a\, \rho(N) \\
&\quad\:\: + \frac{1}{2h} \int_N d|\rho-\nu|(\x) \int_N d|\rho-\nu|(\y)\: \L(\x,\y) \:.
\end{split}
\eeq
We vary~$\rho$ within the class of all regular Borel measures on~$N$ satisfying the constraints~\eqref{rhoprop}.
We note that the function~$\phi$ will be referred to as the {\em{external potential}}
(similar as considered in the so-called inner variational principle in~\cite{cauchy}).
It gives the freedom to modify the variational principle to the effect that the measure~$\rho$ is
favored to be supported in the region where~$\phi$ is small.
This will be needed in Section~\ref{secscalar} in order  to construct a flow of measures emanating 
from a single point.

We define the positive and negative parts of the signed measure~$\rho-\nu$ as usual by
\[ (\rho-\nu)_\pm := \frac{1}{2}\:\Big( |\rho-\nu| \pm (\rho-\nu) \Big) \:. \]
We define the function
\[ 
\epsilon \::\: N \rightarrow \{\pm 1\} \quad \text{by} \quad \epsilon(x) = \left\{
\begin{array}{cl} -1 & \text{if~$x \in \supp(\rho-\nu)_-$} \\
1 & \text{otherwise} \:. \end{array} \right. \]
We introduce the function
\beq \label{elldef}
\ell_\rho(\x) = -\int_N \L(\x,\y)\: d\nu(\y)  + \phi(x)+ a + \frac{1}{h}\:\epsilon(\x) \int_N \L(\x,\y)\: d|\rho-\nu|(\y)\:.
\eeq
Finally, we define the following subsets of~$N$,
\begin{align*}
N_O &:= \supp \rho  && \text{\em{occupied set}} \\
N_U &:= \supp (\mu - \rho) && \text{\em{unsaturated set}} \\
N_B &:= N_O \cap N_U && \text{\em{boundary set}}\:.
\end{align*}

\begin{Prp} {\bf{(Existence of Minimizers and EL Equations)}} \label{prpexist} For any~$a,h  > 0$
and~$\phi \in C^0_0(N, \R)$,
there are solutions of the variational principle where we minimize the action~\eqref{Shdef} varying
within the class of all regular Borel measures on~$N$ satisfying the constraints~\eqref{rhoprop}.
Moreover, every minimizing measure~$\rho$ satisfies the EL equations
\beq \label{ELnc}
\sup_{N_O} \ell_\rho \leq 0 \leq \inf_{N_U} \ell_\rho \:.
\eeq
\end{Prp}
\Proof For~$K$ as in~\eqref{nuprop} we choose the compact set~$K' \subset N$
which contains the support of~$\phi$ and satisfies the condition~\eqref{compactrange}.
Then the action~\eqref{Shdef} can be rewritten as
\begin{align*}
\Sact^h(\rho) &= -\int_{K'} d\rho(\x) \int_N d\nu(\y)\: \L(\x,\y) + \int_{K'} \phi\: d\rho + a\, \rho(N) \\
&\quad\:\: + \frac{1}{2h} \int_N d|\rho-\nu|(\x) \int_N d|\rho-\nu|(\y)\: \L(\x,\y)  \\
&\geq -\int_{K'} d\rho(\x) \int_N d\nu(\y)\: \L(\x,\y) + \int_{K'} \phi\: d\rho + a\, \rho(K') \\
&\quad\:\: + \frac{1}{2h} \int_{K'} d|\rho-\nu|(\x) \int_{K'} d|\rho-\nu|(\y)\: \L(\x,\y) 
= \Sact^h\big(\chi_{K'}\, \rho \big) \:.
\end{align*}
 We let~$(\mu_n)_{n \in \N}$ be a minimizing sequence. Then, according to the previous estimate,
 the sequence~$\chi_{K'} \mu_n$ is also a minimizing sequence.
 Since all the measures are supported inside the compact set~$K'$, a subsequence
 converges due to standard compactness arguments for measures on compact
topological spaces (for details see for example~\cite[Section~3.2]{noncompact}).
Moreover, by continuity the limit measure is a desired minimizer.

In order to prove the EL equations~\eqref{ELnc}, assume conversely
that~$\sup_{N_O} \ell_\rho > 0$ or $0 > \inf_{N_U} \ell_\rho$.
According to the first condition in~\eqref{rhoprop}, we
may form the Radon-Nikodym derivative of~$\rho$ with respect to~$\mu$,
\[ d\rho = f\, d\mu \qquad \text{with~$f : N \rightarrow [0,1]$ measurable} \:. \]

Suppose that~$\sup_{N_O} \ell_\rho > 0$. Then there is a set~$\Omega_O \subset N_O$
with~$\rho(\Omega_O) > 0$ and~$\ell_\rho|_{\Omega_O} > 0$.
We consider the variation
\beq \label{rhovary}
\tilde\rho_\tau = \rho - \tau \chi_{\Omega_O} \rho \qquad \text{for} \qquad
0 \leq \tau \ll 1\:.
\eeq
This variation satisfies the constraint. Moreover, a direct computation shows that the
action~$\Sact^h(\tilde\rho_\tau)$ decreases along the path, in contradiction to the fact that~$\rho$
is a minimizer.

Similarly, if~$\inf_{N_U} \ell_\rho < 0$, we choose a compact subset~$\Omega_U \subset N_U$
with~$\mu(\Omega_U) >0$ and~$\ell_\rho|_{\Omega_U} < 0$. We consider the variation
\beq \label{rhovary2}
\tilde\rho_\tau = \rho + \tau\: \chi_{\Omega_U} \:\mu \qquad \text{for} \qquad
0 \leq \tau \ll 1\:.
\eeq
Again, this variation satisfies the constraint and decreases the action. This is again a contradiction,
concluding the proof.
\QED

We note for clarity that the EL equations~\eqref{ELnc} imply that the function~$\ell_\rho$
is constant on the boundary set,
\[ \ell_\rho |_{N_B} = 0 \:. \]
In the case that~$N$ has a smooth manifold structure and assuming
that zero is a regular value of the function~$\ell_\rho$,
it follows that the measure~$\rho$ is of the form~\eqref{nudef} with
\[ \Omega := \ell_\rho^{-1} \big( (-\infty, 0) \big) \:. \]
With this in mind, in the examples in Sections~\ref{seceuclid} and~\ref{secriemann}, we will restrict our attention to measures of the form~\eqref{nudef}.
However, we point out that, at present, there is no argument which explains why zero
should be a regular value of~$\ell_\rho$. A first step in this direction would be to derive
upper bounds for the total volume~$\mu(N_B)$ of the boundary set.

We finally remark that the set~$\supp(\rho-\nu)_- \cap \supp(\rho-\nu)_+$
could be non-zero and could even have positive measure (this is because the
total variation measure is constructed using general Borel sets, whereas the supports of
measures are by definition closed). For~$x$ in this set, the sign of the
last summand in~\eqref{elldef} can be flipped. Using this freedom, at~$x$ one gets
stronger EL equations. We do not enter the details because this estimate will not be
needed here.

\subsection{A Few A-Priori Estimates}
We now derive a few a-priori estimates which will be needed for the minimizing
movement method in Section~\ref{seclip}.
In preparation, we give an estimate which was first derived similarly
in~\cite[Lemma~2.12]{continuum}.
\begin{Lemma} \label{lemmapartition}
Let~$L \subset N$ be compact. Then there is a constant~$\varepsilon=\varepsilon(L)>0$
such that for any regular Borel measure~$\rho$ on~$L$,
\[ \int_L d\rho(\x) \int_L d\rho(\y)\: \L(\x,\y) \geq \varepsilon\: \rho(L)^2 \:. \]
\end{Lemma}
\Proof Using that the Lagrangian is continuous and strictly positive on the diagonal (assumptions~(ii) and~(v)),
every~$x \in L$ has a neighborhood~$U$ such that~$\L|_{U \times U} >c$ with~$c>0$.
We cover~$L$ by a finite number~$U_1, \ldots, U_N$ of such neighborhoods, i.e.\
\[ \L|_{U_i \times U_i} \geq c_i > 0 \qquad \text{for all~$i=1,\ldots, N$}\:. \]
Setting~$L_1 = U_1$, $L_2 =U_2 \setminus L_1$, \ldots, $L_i = U_i \setminus (L_1, \ldots, L_{i-1})$
(with $i=1,\ldots, N)$ we can arrange that the~$L_i \subset \Omega_i$ form a partition of~$L$.
Then
\begin{align*}
&\int_L d\rho(\x) \int_L d\rho(\y)\: \L(\x,\y)
= \sum_{i,j=1}^N \int_{L_i} d\rho(\x) \int_{L_j} d\rho(\y)\: \L(\x,\y) \\
&\geq \sum_{i=1}^N \int_{L_i} d\rho(\x) \int_{L_i} d\rho(\y)\: \L(\x,\y) \geq \sum_{i=1}^N c_i\: \rho(L_i)^2 \:.
\end{align*}
On the other hand, the Schwarz inequality gives
\[ \rho(L) = \sum_{i=1}^N \rho(L_i) \leq \sqrt{N}\: \Big( \sum_{i=1}^N \rho(L_i)^2 \Big)^\frac{1}{2} \:. \]
Combining these inequalities, we obtain
\[ \int_L d\rho(\x) \int_L d\rho(\y)\: \L(\x,\y) \geq \Big( \min_{k=1,\dots, N} c_k \Big) \;
\frac{1}{N}\: \rho(L)^2 \:, \]
giving the desired estimate with~$\varepsilon = \min_k c_k/N$.
\QED

\begin{Lemma} \label{lemmalip} For any compact subset~$L \subset N$, there is a constant~$\varepsilon(L)>0$ such that
every minimizer~$\rho$ of the action~\eqref{Shdef} satisfies the estimate
\beq \label{totvares}
\|\rho - \nu\|_L \leq Ch \qquad \text{with} \qquad
C := \frac{\s+|a| + \|\phi\|_0}{\varepsilon(L)} \:,
\eeq
where~$\|\rho - \nu\|_L := |\rho - \nu|(L)$ is the total variation norm on~$L$.
\end{Lemma}
\Proof
We integrate the EL equations~\eqref{elldef} over~$L$ with respect to
the measure~$\rho-\nu$. Using that
\[ \supp (\rho - \nu)_+ \subset N_O \qquad \text{and} \qquad
\supp (\rho - \nu)_- \subset N_U \:, \]
we obtain
\begin{align*}
0 &\geq -\int_L d(\rho - \nu)(\x) \int_N d\nu(\y)\: \L(\x,\y) + \int_L \phi\: d(\rho-\nu) \\
&\quad\: + \frac{1}{h}\:\int_L d|\rho - \nu|(\x) \int_N d|\rho - \nu|(\y)\: \L(\x,\y) + |a|
\, (\rho - \nu)(L) 
\end{align*}
and thus
\beq \label{inter}
\int_L d|\rho - \nu|(\x) \int_N d|\rho - \nu|(\y)\: \L(\x,\y)
\leq h\: \|\rho - \nu\|_L \: \big( \s + |a|+ \|\phi\|_0 \big) \:.
\eeq

Applying Lemma~\ref{lemmapartition}, we obtain
\begin{align*}
 \int_L d\rho(\x) \int_N d\rho(\y)\: \L(\x,\y) \geq 
 \int_L d\rho(\x) \int_L d\rho(\y)\: \L(\x,\y) \geq \varepsilon(L)\, \rho(L)^2
\end{align*}
with~$\varepsilon(L)>0$. Using this inequality for~$\rho = |\rho-\nu|$, we conclude that
\[ \varepsilon(L)\: \|\rho - \nu\|_L^2
\leq h\: \|\rho - \nu\|_L \: \big( \s + |a|+ \|\phi\|_0 \big) \:, \]
giving the result.
\QED

It is important to note that the above estimate~\eqref{totvares} is {\em{not}} uniform in the
size of the compact set~$L$ (as is obvious because~$\varepsilon(L)$ appears in the denominator).
But, at least, one can get a uniform estimate by taking the limit~$L \nearrow N$ in~\eqref{inter}.
We now work out this estimate as a separate proposition.
In order to give a different perspective, for the derivation we work directly with the action rather than 
with the EL equations.
\begin{Prp} Every minimizing measure~$\rho$ satisfies the inequality
\[ \int_N d|\rho-\nu|(\x) \int_N d|\rho-\nu|(\y)\: \L(\x,\y) \leq 2h \:\|\rho - \nu\|\: \big( \s + |a|+ \|\phi\|_0 \big)\:. \]
\end{Prp}
\Proof Since~$\rho$ is a minimizer, we know that
\begin{align*}
S^h(\nu) &= 
 -\int_N d\nu(\x) \int_N d\nu(\y)\: \L(\x,\y) +  \int_N \phi\: d\nu + a\, \nu(N)\\
&\geq S^h(\rho) = -\int_N d\rho(\x) \int_N d\nu(\y)\: \L(\x,\y) + \int_N \phi\: d\rho + a\, \rho(N) \\
&\qquad\qquad\quad + \frac{1}{2h} \int_N d|\rho-\nu|(\x) \int_N d|\rho-\nu|(\y)\: \L(\x,\y) \:,
\end{align*}
giving rise to the inequality
\begin{align}
&\int_N d|\rho-\nu|(\x) \int_N d|\rho-\nu|(\y)\: \L(\x,\y) \notag \\
&\leq 2h \big( 
\int_N d(\rho - \nu)(\x) \int_N d\nu(\y)\: \L(\x,\y) - \int_N \phi\: d(\rho-\nu) - a\, (\rho-\nu)(N) \Big) \:.
\label{voles}
\end{align}
Using the EL equations~\eqref{EL} and~\eqref{ldef} together with the non-negativity of the Lagrangian gives the result.
\QED

\subsection{Prescribing the Volume Change}
We refer to the quantity~$(\rho-\nu)(N) \in \R$ as the {\em{volume change}}.
It is implicitly determined by the parameter~$a$ in~\eqref{Shdef}. Qualitatively speaking,
if~$a$ is chosen smaller, then the volume change increases.
In order to get more quantitative control of the volume change, it is useful to prescribe it by imposing the {\em{volume constraint}}
\beq \label{volconstraint}
(\rho-\nu)(N) = h\, b
\eeq
with a new parameter~$b>0$. If we minimize under this constraint, the compactness arguments
of Proposition~\ref{prpexist} still apply. However, the argument which made it possible
to restrict attention to a compact subset~$K' \subset N$ needs to be modified.
This also makes it necessary to impose an additional condition.
\begin{Prp} \label{prpexvol}
Given~$a,b,h  > 0$ and~$\phi \in C^0_0(N, \R)$, assume that the set
\beq \label{Bdef}
B := \Big\{ x \in N \:\Big|\: \int_N \L(\x,\y)\: d\nu(\y) - \phi(x) > b\, \sup_{x,y \in N} \L(x,y)
\Big\}
\eeq
satisfies the inequality
\beq \label{exvolcond}
(\mu - \nu)(B) > b h\:. 
\eeq
Then there are solutions of the variational principle where we
minimize the action~\eqref{Shdef} in the class of all regular Borel measures on~$N$ satisfying the constraints~\eqref{rhoprop}
as well as the volume constraint~\eqref{volconstraint}.
\end{Prp} \noindent
We note that the condition~\eqref{exvolcond} can be
satisfied by choosing~$\phi$ to be negative and by making~$b$ sufficiently small.
\Proof[Proof of Proposition~\ref{prpexvol}.]
We choose the compact set~$K'$ as in the proof of Proposition~\ref{prpexist}.
Let~$\rho$ be a regular Borel measures on~$N$ which satisfy the volume 
constraint~\eqref{volconstraint}. We want to show that the action becomes smaller
if the part of the measure~$\rho$ supported outside~$K'$ is moved inside~$K'$.
This is ensured by the condition~\eqref{exvolcond}, if one keeps in mind that
the left side of the inequality~\eqref{Bdef} tells us how the action is decreased
per volume, whereas the right side gives the maximal increase per volume.
\QED

In order to work out the corresponding EL equations, we consider in modification of
both~\eqref{rhovary} and~\eqref{rhovary2} the variations 
\[ 
\tilde\rho_\tau = \rho - \frac{\tau}{\rho(\Omega_O)}\: \chi_{\Omega_O} \,\rho + \frac{\tau}{\mu(\Omega_U)}\, \chi_{\Omega_U} \:\mu
\qquad \text{for} \qquad 0 \leq \tau \ll 1\:. \]
This variation is admissible and satisfies the volume constraint. A direct computation gives
\beq \label{firstvary}
0 \leq \delta \Sact^h
= - \frac{1}{\rho(\Omega_O)} \int_{\Omega_O} \ell_\rho\: d\rho
+ \frac{1}{\mu(\Omega_U)}\: \int_{\Omega_U} \ell_\rho\: d\mu
\eeq
(note that here the value of the parameter~$a$ in~\eqref{elldef} is irrelevant, because~$a$ drops
out of~\eqref{firstvary}). We conclude that the EL equation~\eqref{ELnc} again holds if we choose
\beq \label{aval}
a = \inf_{\x \in N_O} \bigg( \int_N \L(\x,\y)\: d\nu(\y)  - \phi(x) - \frac{1}{h}\:\epsilon(\x) \int_N \L(\x,\y)\: d|\rho-\nu|(\y)
\bigg) \:.
\eeq
In this way, we have recovered the parameter~$a$ in the action~\eqref{Shdef} as the Lagrange
parameter of the volume constraint~\eqref{volconstraint}. The following estimates hold independently
of whether~$a$ is given a-priori or is determined a-posteriori by~\eqref{aval}.

We finally derive an a-priori bound for the parameter~$a$.
\begin{Lemma} \label{lemmaapriori}
The parameter~$a$ as introduced by~\eqref{aval} is bounded by
\[ 0 < a \leq \s + \|\phi\|_0 \:. \]
\end{Lemma}
\Proof The condition~\eqref{exvolcond} ensures that~$a>0$.
In order to derive the upper bound for~$a$, we employ
the volume constraint~\eqref{volconstraint} in the estimate~\eqref{voles} to obtain
\begin{align*}
&\int_N d|\rho-\nu|(\x) \int_N d|\rho-\nu|(\y)\: \L(\x,\y) \notag \\
&\leq 2h \big( 
\int_N d(\rho - \nu)(\x) \int_N d\nu(\y)\: \L(\x,\y) - \int_N \phi\: d(\rho-\nu) - a b h\big) \:.
\end{align*}
Applying Lemma~\ref{lemmapartition}, we get
\begin{align*}
\varepsilon\, b^2 \: \|\rho- \nu\|^2
&\leq 2h \big( bh \s + bh\, \|\phi\|_0 - a b h\big) \:,
\end{align*}
giving the result.
\QED
In view of this lemma, in the following constructions we may always assume that the
parameter~$a$ is uniformly bounded.

\subsection{Construction of Lipschitz Curves of Measures} \label{seclip}
Inspired by the method of minimizing movements~\cite{degiorgi-minmov, ambrosio-minmov, braides-minmov},
we now explain how to construct a curve~$\rho(t)$ of measures which satisfies
a generalization of the flow equation~\eqref{flowsmooth} to the non-smooth setting.
In order to have large flexibility to adapt the flow to different applications, we consider two
different situations: In the first situation, the parameter~$a$ is chosen as
an arbitrary continuous functions
of time, the area and volume. In the second case, on the other hand, we prescribe the volume change
as an arbitrary function of time. Here is our main result.
\begin{Thm} \label{thmlip}
Given~$T \in \R^+$ (the final time) and an initial measure~$\nu$ of the form~\eqref{nuprop}, consider one of the two following situations:
\bitem
\item[\rm{(i)}] We are given a continuous function $a : [0,T] \times \R^+_0 \times \R^+_0 \rightarrow \R^+$.
\item[\rm{(ii)}] We are given a continuous function $b : [0,T] \rightarrow \R^+$
satisfying the constraint
\beq
\int_0^T b(t)\: dt < (\mu - \nu) \Big( \big\{ x \in N \:\big|\: \phi(x) < 0 \big\} \Big)
  \:. \label{exvolcond2}
\eeq
\eitem
Then there is a Lipschitz continuous curve~$\rho(t)$ with~$t \in [0,T]$ and~$\rho(0)=\nu$,
obtained by minimizing movements from the action~\eqref{Shdef}
with
\[ a=a \big(t,A(t), V(t) \big)\quad \text{(case~{\rm{(i)}})} \qquad \text{or} \qquad \frac{d}{dt} V(t) = b(t) \quad\text{(case~{\rm{(ii)}})}\:, \]
where~$A$ and~$V$ denote the area and volume of~$\rho(t)$, respectively (see Definition~\ref{defarea}).
\end{Thm}

The remainder of this section is devoted to the proof of this theorem.
Given~$h>0$ sufficiently small, we proceed inductively. The starting point is a measure~$\nu=:\rho^h_0$ having the
property~\eqref{nuprop}. For the induction step, given a measure~$\nu=\rho^h_n$,
in case (i) we choose~$a(nh, A(\nu), V(\nu))$ and minimize without volume constraint,
whereas in case~(ii) we choose~$b=b(nh)$, minimize with prescribed volume change~\eqref{volconstraint}
and choose~$a$ according to~\eqref{aval}.
In the latter case, the condition~\eqref{exvolcond2}
ensures that there is a measure~$\rho$ with~$(\rho-\nu)(N)=bh$
which satisfies the condition~\eqref{Bdef} in Proposition~\ref{prpexvol}
for sufficiently small~$h$. Denoting the minimizer by~$\rho$, we set
\[ \rho^h_{n+1} := \rho \:. \]
Iterating this procedure, we obtain a sequence of measures~$(\rho^h_n)_{n \in \N_0}$.
Out of this sequence, we form a piecewise constant curve
\[ 
\rho^h : \R \rightarrow \meas \:,\qquad 
\rho^h(t) := \rho^h_{\lfloor\frac{t}{h}\rfloor} \]
(where~$\lfloor \cdot \rfloor$ denotes the Gau{\ss} bracket).
Combining the estimate in Lemma~\ref{lemmalip} with the triangle inequality immediately gives the 
following Lipschitz-type estimate in any compact set.
\begin{Prp} \label{prpapriori} For any compact subset~$L \subset N$, there is a constant~$C(L)>0$ such that
every minimizer~$\rho$ of the action~\eqref{Shdef} satisfies the estimate
\beq \label{totvares2}
\big\| \rho^h(t) - \rho^h(t') \big\|_L \leq C\, \big( |t-t'| + 2 h\big) \:.
\eeq
\end{Prp}

The final step is to take the limit~$h \searrow 0$ for suitable subsequences.
\begin{Prp} There is a sequence~$(h_j)_{j \in \N}$ with~$h_j \searrow 0$ such that the piecewise
constant curves converge,
\[ \rho^{h_j}(t) \rightarrow \rho(t) \qquad \text{in~$\meas$ for all~$t \in \R$}\:. \]
The resulting curve~$\rho : \R \rightarrow \meas$ is Lipschitz continuous in the sense that
for any compact subset~$L \subset N$ there is a constant~$C(L)>0$ such that
\[ 
\big\| \rho(t) - \rho(t') \big\|_L \leq C\, |t-t'| \:. \]
\end{Prp}
\Proof We choose a compact set~$L \subset N$ and a sequence~$(t_n)_{n \in \N}$
which is dense in~$\R$ (for example an enumeration of the rationals).
According to the constraints~\eqref{rhoprop}, all the measures constructed in Proposition~\ref{prpexist}
are uniformly bounded on~$L$, because
\[ \rho(L) \leq \nu(L) < \infty \:. \]
We now proceed as follows.
We let~$(h_j)_{j \in \N}$ be a sequence with~$h_j \searrow 0$.
For any~$n \in \N$, the sequence of measures~$(\rho^j_n)_{j \in \N}$ with~$\rho^j_n := \rho^{h_j}(t_n)$
has a convergent subsequence. We choose a subsequence~$h_{j_\ell}$
such that the measures~$(\rho^{j_\ell}_1)$ converge, then
a subsequence of this subsequence such that~$(\rho^{j_\cdot}_2)$ converges, and so on.
Denoting the resulting diagonal sequence again by~$(h_j)_{j \in \N}$, it follows that
\beq \label{denseconverge}
\rho^{h_j}(t_n) \overset{j \rightarrow \infty}{\longrightarrow} \rho(t_n) \qquad \text{for all~$n \in \N$} \:,
\eeq
with convergence in~$\meas(L)$.

Let us show that the sequence of measures~$\rho^{(h_j)}(t)$ converges even for every~$t \in \R$.
Given~$t$ and~$\varepsilon>0$, we choose~$n$ such that~$|t_n-t| < \varepsilon/(4C)$
(with the constant~$C$ as in~\eqref{totvares2}). Then the a-priori estimate in Proposition~\ref{prpapriori} yields
\begin{align*}
&\| \rho^{(h_j)}(t) - \rho^{(h_k)}(t) \|_L \\
&\leq  \| \rho^{(h_j)}(t) - \rho^{(h_j)}(t_n) \|_L + \| \rho^{(h_j)}(t_n) - \rho^{(h_k)}(t_n) \|_L
+ \| \rho^{(h_k)}(t_n) - \rho^{(h_k)}(t) \|_L \\
&\leq  \| \rho^{(h_j)}(t_n) - \rho^{(h_k)}(t_n) \|_L + 2\, C\, \big( |t_n-t| + 2 h_j+2 h_k \big) \\
&\leq  \| \rho^{(h_j)}(t_n) - \rho^{(h_k)}(t_n) \|_L + \frac{\varepsilon}{2} +  4\, C\, \big( h_j+h_k \big) 
\xrightarrow{j,k \rightarrow \infty } 0 \:,
\end{align*}
where in the last step we used~\eqref{denseconverge} as well as the fact that the sequence~$(h_j)$ tends to zero.

In order to show Lipschitz continuity of the limit curves of measures, we consider the estimate~\eqref{totvares2}
for the measures~$\rho^{h_j}$,
\[ \big\| \rho^{h_j}(t) - \rho^{h_j}(t') \big\|_L \leq C\, \big( |t-t'| + 2 h_j \big) \:. \]
Taking the limit~$j \rightarrow \infty$ gives the result.
\QED
This also concludes the proof of Theorem~\ref{thmlip}.

\begin{Remark} {\em{
One could also penalize with the Wasserstein distance (similar as done in~\cite{minmov}).
However, this would make it necessary to introduce a distance function~$d : \G \times \G \rightarrow \R$.
The method presented here has the advantage that one works purely with the
Lagrangian~$\L(\x,\y)$. The only drawback is that we can control the flow only locally
(i.e.\ in any compact set~$L \subset N$).
}} \QEDrem
\end{Remark}

\section{Example: The Isoperimetric Flow in Euclidean Space} \label{seceuclid}
We now return to the setting of Euclidean space already introduced in the motivation
in Section~\ref{seciso}. Our goal is to perform a systematic asymptotic expansion
of our area functional for small~$\delta$.

\subsection{Rewriting Surface Layer Integrals as Surface Integrals}
The area functional introduced in~\eqref{Adef} is a surface layer integral.
As already mentioned in~\eqref{Aasy}, this area functional indeed agrees with the
two-dimensional area, up to a prefactor and up to higher order corrections
in the range~$\delta$ of the Lagrangian. In order to compute the corrections, we
use a method first introduced in~\cite[Section~5.1]{matter} which make it possible
to rewrite surface layer integrals as surface integrals. For the sake of self-consistency,
we here review this method and adapt it to our setting.

As in Section~\ref{seciso} we assume that~$\G=\R^3$ is the three-dimensional Euclidean space,
Moreover, we again let~$\mu$ be the Lebesgue measure on~$\G$,
\[ 
d\mu(\x) = d^3\x \qquad \text{and thus} \qquad N \simeq \R^3 \:. \]
Finally, we let~$\Omega \subset N$ be an open subset with smooth boundary~$\partial \Omega$.
It is useful to rewrite the area functional~\eqref{Adef} as
 \[
A(\Omega) = \int_{\Omega} d\mu(x) \int_{N \backslash \Omega} \mathcal{L}(x, y) \:d\mu(y) =
\frac{1}{2} \int_{\Omega} \:d\mu(x) \int_{N } \mathcal{K}(x, y) \:d\mu(y)
\]
with  
\beq \label{Kdef}
\mathcal{K}(x, y) :=
\mathcal{L}(x, y)\: \big( \epsilon_\Omega(\x) - \epsilon_\Omega(y) \big) \:,
\eeq
where~$\varepsilon_\Omega$ is the step function
\[ \epsilon_\Omega(x) = \left\{ \begin{array}{cl}
	1 & \text{if } x \in \Omega \\
	-1 & \text{if } x \not \in \Omega\:.
\end{array} \right. \]

Given a parameter~$s \in [0,1]$, we introduce the new variables~$(\zeta, \xi)$ in~$\R^3$ by
\begin{align*}
	\zeta &= s \x + (1-s)\, \y \:,\hspace*{-3cm} &\xi  &= \y - \x \:. \\
	\x &= \zeta - (1-s) \xi \:,\hspace*{-3cm} & \y &= \zeta + s \xi \:.
\end{align*}
Note that~$\zeta$ is a convex combination of~$\x$ and~$\y$. In this way, varying the parameter~$s$
from zero to one, one can continuously deform these variables, interchanging the roles of~$\x$ and~$\y$. We introduce the function
\[ 
\varphi(s, \zeta) := \int_{\mathbb{R}^3} \mathcal{K} \big( \zeta - (1-s) \xi, \zeta + s \xi \big)\:  d^3\xi \:. \]
Note that, in the case~$s=0$, the variable~$\zeta$ coincides with~$y$,
whereas in the case~$s=1$, it coincides with~$x$. This makes it possible
to write the area functional as
\[ 
A(\Omega) = \frac{1}{2} \int_{\Omega} \varphi(1, \zeta)\: d^3 \zeta
= \frac{1}{4} \int_{\Omega} \Big( \varphi(1, \zeta) - \varphi(0, \zeta) \Big)\: d^3 \zeta \:. \]
Due to the step functions in~\eqref{Kdef}, the kernel~$\mathscr{K}$ is not
differentiable. In order to bypass this problem, we regularize on a length scale~$\varepsilon$
(for example by taking the convolution with a mollifier) and denote the resulting
mollified kernel by~$\mathscr{K}^\varepsilon$. Then the
$s$-derivative of the resulting integral~$\varphi^\varepsilon$ can be computed as follows,
\begin{align*}
		\frac{d}{ds} \varphi^\varepsilon(s, \zeta) &= \frac{d}{ds} \int_N \mathcal{K}^\varepsilon\big( \zeta - (1-s) \xi, \zeta + s \xi \big)\: d^3\xi \\&= \int_N \frac{d}{ds} \mathcal{K}^\varepsilon\big( \zeta - (1-s) \xi, \zeta + s \xi \big)\: d^3\xi\\
		&= \int_N \sum_{\alpha=1}^3 \xi^\alpha \: \frac{\partial}{\partial \zeta^\alpha} \mathcal{K}^\varepsilon\big( \zeta - (1-s) \xi, \zeta + s \xi \big)\: d^3\xi \:.
	\end{align*}
We thus obtain
\[ \varphi^\varepsilon(1, \zeta) - \varphi^\varepsilon(0, \zeta) = \frac{\partial}{\partial \zeta^\alpha}\: (B^\varepsilon)^\alpha(\zeta) \]
with
\[ B^\varepsilon(\zeta) := 
\int_N  d^3\xi\:\xi \int_0^1 ds  \:\mathcal{K}^\varepsilon\big( \zeta - (1-s) \xi, \zeta + s \xi \big) \:. \]
Integrating over~$\Omega$, we can apply the Gau{\ss} divergence theorem to obtain
\[ \int_\Omega \big( \varphi^\varepsilon(1, \zeta) - \varphi^\varepsilon(0, \zeta) \big)\:
d\mu(\zeta) = \int_{\partial \Omega} \la B^\varepsilon, \nu \ra\: d\mu_{\partial \Omega} \:, \]
where~$\nu$ is the outer normal on~$\partial \Omega$.
Since this formula does not involve any derivatives, we may take the limit~$\varepsilon
\rightarrow 0$ to obtain
\[ \int_\Omega \big( \varphi(1, \zeta) - \varphi(0, \zeta) \big)\:d\mu(\zeta) 
= \int_{\partial \Omega} \la B, \nu \ra\: d\mu_{\partial \Omega} \]
with
\[ B(\zeta) := 
\int_N  d^3\xi\:\xi \int_0^1 ds  \:\mathcal{K}\big( \zeta - (1-s) \xi, \zeta + s \xi \big) \:. \]

We now choose the Lagrangian again in the form~\eqref{Leuclid}
with~$\eta$ as in~\eqref{etaprop}. Then
\begin{align*}
B(\zeta) &=
\int_N  \xi\: \eta\Big( \frac{\|\xi\|}{\delta} \Big) \bigg( \int_0^1 
\Big( \epsilon_\Omega\big(\zeta - (1-s) \xi \big) - \epsilon_\Omega \big( \zeta + s \xi \big) \Big)\: ds
\bigg)\:d^3\xi \\
&= \delta^4 \int_N \xi \: \eta \big( \|\xi\| \big)\:
\bigg( \int_0^1  \Big( \epsilon_\Omega\big(  \zeta - (1-s) \:\delta\xi \big)
- \epsilon_\Omega \big( \zeta + s \:\delta\, \xi \big) \Big) \:ds \bigg)\: d^3\xi \:.
\end{align*}

Our findings can be summarized as follows.
\begin{Lemma} \label{lemmadivphi} Under the above assumptions, the
area functional~\eqref{Adef} can be written as a surface integral
\[ A(\Omega) =\frac 14  \int_{\partial \Omega} \la B, \nu \ra\: d\mu_{\partial \Omega} \:, \]
where the vector field~$B$ is given by
\beq \label{Bdef2}
B(\zeta) := \delta^4 \int_N \xi \: \eta \big( \|\xi\| \big)\:
\bigg( \int_0^1  \Big( \epsilon_\Omega\big(  \zeta - (1-s) \:\delta\xi \big)
- \epsilon_\Omega \big( \zeta + s \:\delta\, \xi \big) \:ds \bigg)\: d^3\xi \:.
\eeq
\end{Lemma}

\subsection{Expansion of the Area Functional for Small Range~$\delta$} \label{secdelta}
The goal of this section is to prove the following expansion of our area functional.
\begin{Thm} \label{thmeuclid}
Let~$\Omega \subset N=\R^3$ be open and relative compact with smooth boundary~$\partial \Omega$. Choosing the Lagrangian 
in the form~~\eqref{Leuclid} with~$\eta \in C^\infty((-1,1))$ as in~\eqref{etaprop}, the
area functional~\eqref{Adef} has the following expansion for small~$\delta$,
\[ 4 A(\Omega) = \delta^4\, \s_1\, \mu_2(\partial \Omega)
- \delta^6\, \s_2  \int_{\partial \Omega} \big(  3 H^2 - 4 K \big)\:
d\mu_{\partial \Omega}  + \O\big( \delta^8 \big) \:, \]
where~$H$ and~$K$ are the mean and Gau{\ss} curvatures of~$\partial \Omega$
(with the convention that the mean curvature is the sum of the principal curvatures),
and the constants~$\s_1$ and~$\s_2$ are given by~\eqref{s1def} and
\beq \label{s2def}
\s_2 := \frac{\pi}{6} \int_0^\infty \eta(r)\: r^5\: dr\:.
\eeq
\end{Thm}

The proof is based on the following lemma.
\begin{Lemma} \label{lemmaeuclid} Let~$\Omega \subset N$ be open 
(not necessarily relatively compact) with smooth boundary~$\partial \Omega$.
Then, for any~$\zeta \in \partial \Omega$, the vector field~$B(\zeta)$ in~\eqref{Bdef2}
has the following expansion for small~$\delta$,
\beq \label{Bform}
B(\zeta) = \delta^4\, \s_1\: \nu - \delta^6\, \s_2 \,\big( 3 H^2 - 4 K \big)\: \nu  + \O\big( \delta^8 \big) \:,
\eeq
where~$H$ and~$K$ are the mean and Gau{\ss} curvatures of~$\partial \Omega$
(and~$\nu$ is again the outer normal).
Moreover, the constants~$\s_1$ and~$\s_2$ are given by~\eqref{s1def} and~\eqref{s2def}.
\end{Lemma}
\Proof Given~$\zeta \in \partial \Omega$, we arrange by a translation and rotation that~$\zeta=0$
and that the surface is tangential to the $(x,y)$-plane. Then the set~$\Omega$ can be described locally by the
inequality
\[ z < f(x,y) \]
with a suitable smooth function~$f(x,y)$ (thus $\partial \Omega$ is represented locally by the
graph~$z=f(x,y)$). We write~$\xi=(\hat{\xi},z)$ and~$\hat{\xi}=(x,y)$.

In preparation, we consider the case~$f=0$, so that~$\Omega$ is the half plane~$z<0$. In this case,
\beq \label{epssimp}
\epsilon_\Omega\big(- (1-s) \:\delta\,\xi \big)
- \epsilon_\Omega \big( s \:\delta\, \xi \big) = 2 \, \epsilon(z) \:,
\eeq
and thus
\[ B^\alpha(\zeta) = 2 \delta^4 \int_{\R^3} \xi^\alpha \: \eta \big( \|\xi\| \big)\: \epsilon(z)\: d^3 \xi \:. \]
This integral vanishes by symmetry unless~$\alpha=3$. We thus obtain
\[ B = 2 \:\nu\: \delta^4
\int_{\R^3} |z| \: \eta \big( \|\xi\| \big) \: d^3 \xi \:. \]

We next consider the case that $f(x,y)$ is a homogeneous polynomial of degree two. Then
\[ \epsilon_\Omega \big( s \:\delta\, \xi \big) 
=  \epsilon \Big( f\big(s \,\delta\, \hat{\xi}\big) - s \,\delta\, z \big) \Big) \]
(where~$\epsilon$ denotes the sign function). 
We first consider the case that~$f(\hat{\xi}) \neq 0$.
Then, using that~$f$ is homogeneous
of degree two and that~$s\geq0, \delta > 0$, we obtain
\[ \epsilon_\Omega \big( s \:\delta\, \xi \big) =
\epsilon \Big( s^2\, \delta^2\, f(\hat{\xi}) - s \,\delta\, z \Big)
= \epsilon \big( s\, \delta\, f(\hat{\xi}) - z  \big)
= \epsilon \big( f(\hat{\xi}) \big) \: \epsilon \Big( s - \frac{z}{\delta\,f(\hat{\xi})} \Big) \:. \]
Similarly,
\begin{align*}
\epsilon_\Omega \big( -(1-s) \:\delta\, \xi \big) &=
\epsilon \Big( (1-s)^2\, \delta^2\, f(\hat{\xi}) + (1-s) \,\delta\, z \Big) \\
&= \epsilon \big( f(\hat{\xi}) \big) \: \epsilon \Big( (1-s) + \frac{z}{\delta\,f(\hat{\xi})} \Big) 
= -\epsilon \big( f(\hat{\xi}) \big) \: \epsilon \Big( s - 1 - \frac{z}{\delta\,f(\hat{\xi})} \Big)  \:.
\end{align*}
Combining these formulas, we obtain
\[ \epsilon_\Omega\big(- (1-s) \:\delta\,\xi \big)
- \epsilon_\Omega \big( s \:\delta\, \xi \big)
=  -2\: \epsilon \big( f(\hat \xi) \big)\: \tilde \epsilon_ q(s) \]
with
\[ \tilde \epsilon_q(s) := \left\{ \begin{array}{cl} 
1 & \text{if~$s > 1+ q$} \\
0 & \text{if~$q \leq s \leq 1+q$} \\
-1 & \text{if~$s < q$} \: .
\end{array} \right. \qquad \text{and} \qquad
q := \frac{z}{\delta f (\hat \xi ) } \:.\]
Using these formulas, we can carry out the $s$-integral in~\eqref{Bdef2},
\begin{align}
&\int_0^1 \Big( \epsilon_\Omega\big( -(1-s) \:\delta\xi \big)
- \epsilon_\Omega \big( s \:\delta\, \xi \big)\Big) \:ds \notag \\
&= 2\,\epsilon \big( f(\hat \xi) \big)\: \times \left\{ \begin{array}{cl} -1 & \text{if~$q<-1$} \\
q  & \text{if~$-1 \leq q \leq 1$} \\
1 & \text{if~$q>1$} \end{array} \right.
= \left\{ \begin{array}{cl}
	-2 & \text{if } z < - \vert\delta  f(\hat \xi) \vert  \\[0.3em]
\displaystyle \frac{2 z}{\delta \,\vert f(\hat \xi)\vert }  & \text{if } |z|  \leq \vert\delta  f(\hat \xi) \vert \\[0.9em]
	2& {\text{if } z > \vert\delta  f(\hat \xi) \vert  } \:. \label{sint}
\end{array} \right.
\end{align}
In the remaining case $f(\hat \xi)  = 0$ we get again~\eqref{epssimp},
so that~\eqref{sint} again holds.
Using this result in~\eqref{Bdef2} gives
\begin{align*}
B &= 2 \:\nu\: \delta^4 \:\bigg(
\int_{\R^3} |z| \: \eta \big( \|\xi\| \big) \: d^3 \xi  + \int_{\R^2} d^2 \hat{\xi}
\int_{- \delta \,|f(\hat \xi)|}^{\delta\,| f(\hat \xi)|} dz\: \eta \big( \|\xi\| \big)\:
\Big( \frac{z^2}{\delta\, |f(\hat{\xi})|} - |z| \Big)  \bigg) \:.
\end{align*}
The first integral gives the first summand in~\eqref{Bform}.
The other integrals can be expanded in powers of~$\delta$,
\begin{align*}
& \int_{\R^2} d^2 \hat{\xi} \int_{- \delta \,|f(\hat \xi)|}^{\delta\,| f(\hat \xi)|} dz\: \eta \big( \|\xi\| \big)\:
\Big( \frac{z^2}{\delta\, |f(\hat{\xi})|} - |z| \Big) \\
&= \int_{\R^2} \eta \big( \|\hat{\xi}\| \big) \bigg( \int_{- \delta \,|f(\hat \xi)|}^{\delta\,| f(\hat \xi)|} dz\: \Big( \frac{z^2}{\delta\, |f(\hat{\xi})|} - |z| \Big) \bigg)\:d^2 \hat{\xi} + \O \big( \delta^4 \big) \\
&= \int_{\R^2} \eta \big( \|\hat{\xi}\| \big) \: \bigg( \frac{1}{\delta\, |f(\hat{\xi})|} 
\: \frac{2}{3}\: \big( \delta\, |f(\hat{\xi})|\big)^3 - \big( \delta\, |f(\hat{\xi})|\big)^2 \bigg)\:d^2 \hat{\xi} + \O \big( \delta^4 \big) \\
&= -\frac{\delta^2}{3} \int_{\R^2} \eta \big( \|\hat{\xi}\| \big) \: |f(\hat{\xi})|^2 \:d^2 \hat{\xi}
+ \O \big( \delta^4 \big) \:.
\end{align*}
By a linear coordinate transformation in the $xy$-plane, we can arrange that
\beq \label{fpoly}
f(x,y) = a x^2 + b y^2
\eeq
with~$a,b \in \R$. Then the integrals can be computed in polar coordinates~$(r, \varphi)$,
\begin{align*}
- &\frac{\delta^2}{3} \int_{\R^2} \eta \big( \|\hat{\xi}\| \big) \: |f(\hat{\xi})|^2 \:d^2 \hat{\xi} \\
&= -\frac{\delta^2}{3} \int_0^\infty dr\:r^5\: \eta(r) \int_0^{2 \pi} d\varphi\:
\big( a^2 \cos^4 \varphi + 2 a b \sin^2 \varphi \cos^2 \varphi +
b^2 \sin^4 \varphi \big) \\
&= -\frac{\delta^2 \pi}{12}\:\big( 3 a^2 + 2 a b + 3 b^2 \big) \int_0^\infty r^5\: \eta(r)\: dr
= -\frac{\delta^2 \pi}{12}
\: \Big(3 H^2- 4 K \Big) \int_0^\infty \eta(r)\: r^5\: dr\:,
\end{align*}
giving the second summand in~\eqref{Bform}.

We finally consider the general case of a smooth function~$f(\hat{\xi})$.
Again by a linear coordinate transformation in the $xy$-plane, we can arrange that~$f$
has the form~\eqref{fpoly} with a cubic error term,
\[ f(\hat{\xi}) =  a x^2 + b y^2 + \O \big( \|\hat{\xi}\|^3 \big) \:. \]
Carrying this error term through the above expansions, one verifies by direct inspection
that the resulting error in~\eqref{Bform} is of the order~$\O(\delta^8)$. This concludes the proof.
\QED
We remark that the next terms in the expansion in powers of~$\delta$ could be computed similarly, giving rise to derivatives of curvature and higher order curvature terms.

Theorem~\ref{thmeuclid} follows immediately by combining Lemmas~\ref{lemmadivphi} and~\ref{lemmaeuclid}.

\subsection{Expansion of the Isoperimetric Flow} \label{secexpiso}
We now explain what the expansion of the area functional in Theorem~\ref{thmeuclid}
tells us about the isoperimetric flow. In order to get a direct connection, we rewrite
the functional~$\Sact^h$ in~\eqref{Shdef} as
\begin{align}
&\Sact^h(\rho) = \frac{1}{2} \int_N d\rho(\x) \int_N d(\mu - \rho)(\y)\: \L(\x,\y) \label{Sh1} \\
&\quad\: + \frac{1}{2} \int_N d(\rho-\nu)(\x) \int_N d(\rho - \nu)(\y)\: \L(\x,\y) -
\frac{1}{2} \int_N d\nu(\x) \int_N d\nu(\y)\: \L(\x,\y) \label{Sh2} \\
&\quad\: + \int_N \phi\: d\rho+ \Big(a- \frac{\s}{2} \Big)\, \rho(N) + \frac{1}{2h} \int_N d|\rho-\nu|(\x) \int_N d|\rho-\nu|(\y)\: \L(\x,\y) \:, \label{Sh3}
\end{align}
where we used that
\begin{align*}
\s \,\rho(N) &= \int_N d\rho(\x) \int_N d\mu(\y)\: \L(\x,\y) \\
&= \int_N d\rho(\x) \int_N d\rho(\y)\: \L(\x,\y)  + \int_N d\rho(\x) \int_N d(\mu - \rho)(\y)\: \L(\x,\y) \:.
\end{align*}
Choosing again~$\rho = \chi_\Omega\, \mu$, the double integral in~\eqref{Sh1}
reduces precisely to the area functional~\eqref{Adef}.
The terms in~\eqref{Sh2} are irrelevant for the minimizing movement flow
(note that the last summand is a constant, whereas the
first summand is by a factor~$h$ smaller than the last summand in~\eqref{Sh3}
which drives the minimizing movement flow).
The summands in~\eqref{Sh3} implement the gradient flow in terms of minimizing movements.
This shows that~$\Sact^h$ indeed describes the gradient flow of our area functional~\eqref{Adef}.
According to Theorem~\ref{thmeuclid}, to leading order four in~$\delta$, our area functional
reduces to a multiple of the two-dimensional area. Consequently, the gradient flow
reduces to the mean curvature flow. To the order~$\sim \delta^6$, the
curvature of~$\partial \Omega$
gives a correction to our area functional via the combination~$3 H^2- 4 K$ of mean and Gau{\ss} 
curvatures. This gives rise to a corresponding correction to the isoperimetric flow
of the form of the gradient of the correction to the area functional.
The even higher orders~$\sim \delta^8, \sim \delta^{10}$, etc., could all be computed and
understood in a similar manner. We point out that all these corrections are combined
in such a way that the isoperimetric flow is well-defined even if the smoothness assumptions
on~$\Omega$ are dropped.

\section{Example: The Isoperimetric Flow in a Riemannian Manifold} \label{secriemann}
We now generalize the previous example to a flow taking place in a
Riemannian manifold of general dimension.
\begin{Thm} \label{thmriemann}
Let~$(N,g)$ be a smooth Riemannian
manifold of dimension~$k>1$. Assume that~$N$ is geodesically complete and
has a positive injectivity radius~$J>1$. We choose the Lagrangian as
\beq \label{Ldelta}
\L(\x,\y) = \eta \Big( \frac{d(x,y)}{\delta} \Big) \:,
\eeq
where~$\eta \in C^\infty((-1,1))$ is again a non-negative test function
of the form~\eqref{etaprop}, and the geodesic distance between~$x$ and~$y$
is denoted by~$d(x,y)$. Then, for small~$\delta \in (0,J)$, the area functional~\eqref{Adef} has the expansion
\begin{align}
4 A(\Omega) &= \delta^{k+1}\, \s_1\, \mu_{k-1}(\partial \Omega) 
- \delta^{k+3} \:\s_2 \int_{\partial \Omega} \Big( H^2 + 2\, \Tr \big( S^2 \big) \Big) \:
d\mu_{\partial \Omega} \label{Ak} \\
&\quad\: -2 \delta^{k+3}\,\s_2 \int_{\partial \Omega} \big( \Ric (\nu, \nu) + 
\text{\rm{scal}} \big)\: d\mu_{\partial \Omega} + \O \big( \delta^{k+4} \big)\:, \label{ARiem}
\end{align}
where~$\nu$ is the outer normal, $\mu_{k-1}$ is the $k-1$-dimensional Hausdorff measure,
$S$ is the Weingarten map,
\beq \label{weingarten}
S \::\: T\partial \Omega \rightarrow T\partial \Omega \:, \qquad
u \mapsto -\nabla_u \nu \:,
\eeq
and~$H=\Tr(S)$ is mean curvature.
Moreover, the Ricci and scalar curvatures of~$N$ are denoted by~$\Ric$ and~$\text{\rm{scal}}$,
respectively. Finally, the parameters~$\s_1$ and~$\s_2$ are given by
\begin{align}
\s_1 &:= \frac{4 \,\pi^\frac{k-1}{2}}{\Gamma \big(\frac{k+1}{2} \big)} 
\int_0^\infty r^k\: \eta(r)\: dr \\
\s_2&:= \frac{\pi^\frac{k-1}{2}}{3 \,\Gamma \big(\frac{k+3}{2} \big)} 
\int_0^\infty r^{k+2}\: \eta(r)\: dr \:.
\end{align}
\end{Thm} \noindent
Before entering the proof, we briefly explain this result.
The contributions in~\eqref{Ak} generalize the result of Theorem~\ref{thmeuclid}
to higher dimension. The contributions in~\eqref{ARiem}, however, 
involve the Ricci curvature of the ambient space~$N$.
Similar as explained in Section~\ref{secexpiso} in Euclidean space, these
curvature corrections also give rise to additional corrections to the isoperimetric flow
of order~$\sim \delta^{k+3}$. Again, the even higher orders~$\sim \delta^{k+4}$, $\sim \delta^{k+5}$, etc., could be computed a straightforward way. For brevity, we shall not enter the details.

The remainder of this section is devoted to the proof of this theorem.
At first sight, the problem seems closely related to Weyl's tube formula
(as developed in detail in the textbook~\cite{gray}).
However, it turns out to be preferable not to work with tubular neighborhoods,
but rather with exponential charts centered at an intermediate point between the two
arguments~$x$ and~$y$ of the Lagrangian.
We note that the curvature expressions in~\eqref{Ak} and~\eqref{ARiem}
have some similarity with those in Weyl's tube formula
(see~\cite[Lemma~4.2 and~eq.~(1.1)]{gray}).
But they are, of course, different because here we do not compute the volume of the
tube, but rather the area as defined by the double integral~\eqref{Adef}.

In more detail, we proceed as follows.
We assume that~$N$ is geodesically complete.
Moreover, we assume that the range of the Lagrangian is smaller than the
injectivity radius. This means that for any two points~$x,y \in N$ with~$\L(x,y) \neq 0$ there
is a unique geodesic~$\gamma(t)$ joining these two points. We parametrize the geodesic
such that
\[ \gamma(0)=x \quad \text{and} \qquad \gamma(1) = y \:. \]
Given a parameter~$s \in [0,1]$, we choose
\beq \label{zetas}
\zeta := \gamma(s) \qquad \text{and set} \qquad \xi_\zeta := \gamma'(s) \in T_\zeta N \:.
\eeq
Then, by standard properties of the exponential map, we can write~$x$ and~$y$ as
\beq \label{xydef}
x = \exp_\zeta \big( -s \xi_\zeta \big) \qquad \text{and} \qquad
y = \exp_\zeta \Big( (1-s) \xi_\zeta \Big) \:.
\eeq
It is convenient to use~$\zeta$ and~$\xi_\zeta$ to describe the pair of points~$(x,y)$.
In order to formalize this method, for any~$\zeta \in N$ we choose~$B_\zeta N \subset T_\zeta N$
as the ball with radius equal to the injectivity radius~$J$ of $N$. We introduce the set
\[ \mathfrak{I} := \big\{ \xi \in T N \:\big|\: \xi_\zeta \in B_\zeta N \big\} \subset TN \:. \]
We also introduce the mapping
\[ \Phi_s : \mathfrak{I} \rightarrow N \times N \:, \qquad \xi_\zeta \mapsto (x,y) \]
with~$x,y$ as in~\eqref{xydef} (the lower index~$s$ clarifies the dependence on the
parameter~$s \in [0,1]$). In order to transform the integral, we have to choose a measure on 
the tangent bundle. The most natural way to do this is to choose on the tangent bundle the Sasaki 
metric, which means that we use the splitting of $T_{\xi _{\zeta}}(TN)$ given by the Levi-Civita 
connection into a horizontal and a vertical part, which are both  isomorphic to $T_\zeta N$ and fix 
the scalar product by setting this splitting  as orthogonal and the metric on each of the summands  
given by the scalar product $g_\zeta$ (for details see~\cite[Section~3]{sasaki}).
Then the transformation of the measures on $TN$  and the product manifold is given  
by a function~$h(x,y)$,
\beq \label{hsdef}
d \big( \mu \times \mu \big)(x,y) = h_s(x,y)\: d\mu(\zeta)\: d^k\xi_ \zeta
\eeq
with~$h_s \in C^\infty(\Phi_s(\mathfrak{I}), \R^+)$. Here $d^k\xi_ \zeta$ denotes the measure on the 
fiber of the tangent bundle given by the Riemannian metric on $N$. The function~$h_s$ is obviously 
symmetric in its two arguments. We compute it in more detail in the next lemma.
\begin{Lemma} \label{lemmariemann} The function~$h_s$ in~\eqref{hsdef}
has the local expansion
\[ h_s(x,y) = 1- \frac{1}{6}\:\Ric|_\zeta (\xi, \xi)+ \mathcal{O}(\|\xi \|^3) \:. \]
\end{Lemma}
\Proof Given~$\zeta \in N$ and~$\xi \in T_\zeta N$, we let~$\gamma(\tau)$
be the geodesic
\[ \gamma(\tau) := \exp_\zeta(\tau \xi) \:. \]
A Jacobi field~$J_{ \xi}$ is a vector field along this geodesic
(i.e.\ $J_\xi(\tau) \in T_{\gamma_\xi( \tau)}N$ for all~$\tau$) which satisfies the
Jacobi equation
\beq \label{Jacobi}
\nabla_\tau^2 J_\xi = R(\dot{\gamma}, J_\xi) \dot{\gamma} \:.
\eeq
The space of solutions is isomorphic to the $2n$-dimensional vector
space of the initial data
\[ \big( J_\xi(0), \dot{J_\xi}(0) \big) \in T_\zeta N \times T_\zeta N\:. \]
In order to calculate the  differential of $\Phi_s$, we use the splitting of $T_{\xi_\zeta} TN $ into a horizontal part and a vertical part. We let~$J _{a,b}$ be the Jacobi field with $J_{a,b} (0) =a$
and~$\dot{J}_{a,b}(0)   = b$ along the geodesic~$\gamma $.
For a vertical vector~$w$, we get
\[ d\Phi_s |_{\xi_\zeta}(w) = \big( J_{0,w}(-s),J_{0, w}(1-s) \big) \]
Likewise, for a horizontal vector $w$,
\[ d\Phi_s|_{\xi_\zeta}(w) = \big( J_{w,0}(-s),J_{w,0}(1-s) \big)\:. \]  
Identifying the tangent spaces along $\gamma$ by parallel transport, we have for the Jacobi fields $J _{a,b}$ along a geodesic $\gamma_v$ the expansion
(for basics see~\cite[Section~5.2]{docarmo-riemannian}; more details can be found in~\cite[Section~9.1]{gray} or~\cite{berger-gauduchon, chavel})
\[ J(t) = a + t\, b - \frac{t^2}{2!}\, (\mathcal{R}_v a)
	- \frac{t^3}{3!} \:\Big( (\mathcal{R}_v b) + \big( (\nabla_v R)(a,v)\,v \big) \Big)
 	+ \mathcal{O}( t^4) \]
with~$\mathcal{R}_v(w) := R(w,v)v$. We thus obtain
\[ d\Phi_s|_\xi =
\begin{pmatrix}
\displaystyle \1 - \frac{s^2}{2}\,\mathcal{R}_\xi & \displaystyle -s \1 + \frac{s^3}{6}\,\mathcal{R}_\xi \\[0.7em]
\displaystyle \1 - \frac{(1-s)^2}{2}\,\mathcal{R}_\xi & \displaystyle (1-s) \1 - \frac{(1-s)^3}{6}\,\mathcal{R}_\xi \end{pmatrix} + \mathcal{O}(|\xi|^3) \:. \]
A direct computation yields
\[ \det d\Phi_s|_\xi = 1- \frac{1}{6}\:\Ric (\xi, \xi)+ \mathcal{O}(|\xi|^3) \:, \]
giving the result (note that the right side is independent of~$s$).
\QED

Similar to~\eqref{Kdef}, we introduce the function
\[ \mathcal{K}(x,y) := \L(x,y)\: \big( \epsilon_\Omega(x) - \epsilon_\Omega(y) \big) \:. \]
Next, given a function~$f \in C^\infty_0(N, \R)$, we define
\[ \varphi(s) = \int_N d\mu(x) \int_N d\mu(y)\: f\big( \zeta(s,x,y) \big)\: \mathcal{K}(x,y) \]
with~$\zeta=\zeta(s,x,y)$ as defined in~\eqref{zetas}.
Then
\begin{align}
\varphi(0) &= \int_N d\mu(x) \int_N d\mu(y)\: f(x)\: \mathcal{K}(x,y) \label{phi0} \\
\varphi(1) &= \int_N d\mu(x) \int_N d\mu(y)\: f(y)\: \mathcal{K}(x,y) = -\varphi(0)
\end{align}
(where in the last step we used that~$\mathcal{K}$ is anti-symmetric in its two arguments).
Hence
\beq \label{phiint}
\varphi(0) = \frac{1}{2}\: \Big( \varphi(0) - \varphi(1) \Big)
= -\frac{1}{2} \int_0^1 \varphi'(s) \: ds \:.
\eeq
The derivative of~$\varphi$ can be written as
\[ \varphi'(s) = \int_N d\mu(x) \int_N d\mu(y)\: D_{\xi_\zeta} f(\zeta)\: \mathcal{K}(x,y) \:. \]
We would like to apply the Gau{\ss} divergence theorem. However, this is not possible
at this stage, because~$\xi$ also depends on~$\zeta$. In order to improve the situation,
we transform the integrals to the parametrization in terms of~$\zeta$ and~$\xi$,
\begin{align*}
\varphi'(s) &= \int_N d\mu(\zeta) \int_{T_\zeta N} d^k \xi_\zeta\:h_s \big( \Phi_s(\zeta, \xi) \big)\:
D_{\xi_\zeta} f(\zeta)\: \mathcal{K}\big(\Phi_s(\zeta, \xi) \big) \\
&= \int_N d\mu(\zeta)\: \partial_j f(\zeta) \int_{T_\zeta N} d^k \xi_\zeta\:\xi_\zeta^j\: 
h_s \big( \Phi_s(\zeta, \xi) \big)\: \mathcal{K}\big(\Phi_s(\zeta, \xi) \big) \:.
\end{align*}
In the last step we pulled the function~$f$ outside the $\zeta$-integral.
This makes it possible to integrate by parts with the Gau{\ss} divergence theorem,
\[ \varphi'(s) = -\int_N d\mu(\zeta) \:f(\zeta)\:\div_\zeta \int_{T_\zeta N} d^k \xi_\zeta\:\xi_\zeta\: 
h_s \big( \Phi_s(\zeta, \xi) \big)\: \mathcal{K}\big(\Phi_s(\zeta, \xi) \big) \:. \]
Using this formula in~\eqref{phiint}, we conclude that
\beq
\varphi(0) = \frac{1}{2} \int_N d\mu(\zeta) \:f(\zeta)\:\div_\zeta 
\bigg( \int_0^1 ds \int_{T_\zeta N} d^k \xi_\zeta\:\xi_\zeta\: 
h_s \big( \Phi_s(\zeta, \xi) \big)\: \mathcal{K}\big(\Phi_s(\zeta, \xi) \big) \bigg) \label{phi0f}
\eeq

We now consider the limiting case where~$f$ converges pointwise
to the characteristic function of~$\Omega$. 
On the left side of~\eqref{phi0f}, we obtain by applying the Lebesgue's dominated convergence theorem to~\eqref{phi0} 
\begin{align*}
\varphi(0) \rightarrow\; & \int_\Omega d\mu(x) \int_N d\mu(y)\: \mathcal{K}(x,y) 
= \int_\Omega d\mu(x) \int_{N \setminus \Omega} d\mu(y)\: \mathcal{K}(x,y) \notag \\
&= 2 \int_\Omega d\mu(x) \int_{N \setminus \Omega} d\mu(y)\: \L(x,y) = 2A(\Omega)
\end{align*}
On the right side of~\eqref{phi0f}, we can also apply Lebesgue's dominated convergence
theorem and the Gau{\ss} divergence theorem once again.
We conclude that the statement of Lemma~\ref{lemmadivphi} again holds for the vector field
\beq \label{Bzetadef}
B(\zeta) :=  \int_0^1 ds \int_{T_\zeta N} d^k \xi_\zeta\:\xi_\zeta\: 
h_s \big( \Phi_s(\zeta, \xi) \big)\: \mathcal{K}\big(\Phi_s(\zeta, \xi) \big)
\eeq

Next, we compute this vector field more explicitly for a Lagrangian of 
the form~\eqref{Ldelta}.
\begin{Lemma} \label{lemma2}
Choosing the Lagrangian as in~\eqref{Ldelta},
the vector field~$B$ in~\eqref{Bzetadef} has the expansion for small~$\delta$
\begin{align}
g_\zeta(B, \nu) &= \delta^{k+1}\, \s_1 - \delta^{k+3}\, \s_2 \big( H^2 + 2\,\vert S
\vert^{2}\big) \label{gnu0} \\
&\quad\: -2 \delta^{k+3}\, \s_2\, \big( \Ric (\nu, \nu)\: + \text{\rm{scal}}  \big) + \O\big( \delta^{k+4} \big) \:, \label{gnu}
\end{align}
where~$\nu$ is the outer normal and the parameters~$\s_{1,2}$ as in the
statement of Theorem~\ref{thmriemann}.
\end{Lemma}
\Proof We consider an exponential chart around~$\zeta \in \partial \Omega$. Then, in 
analogy to~\eqref{Bdef2}, we obtain
\begin{align*}
B(\zeta) &= \delta^{k+1} \int_{T_{\zeta} N} \xi \: \eta \big( \|\xi\| \big)\:
\Big\{ 1- \frac{\delta^2}{6} \Ric (\xi, \xi)
+ \mathcal{O} \big( (\delta\, |\xi|)^3 \big) \Big\}\: \\
&\qquad\qquad \times \bigg( \int_0^1  \Big( \epsilon_\Omega\big(  \zeta - (1-s) \:\delta\,\xi \big)
- \epsilon_\Omega \big( \zeta + s \:\delta\, \xi \big) \:ds \bigg) \: d^k\xi \:,
\end{align*}
where the curly bracket involves the curvature corrections worked out in Lemma~\ref{lemmariemann}.
We now arrange by a rotation that the coordinate plane~$\{ \xi_k=0 \}$ is tangential
to~$\partial \Omega$. We represent~$\partial \Omega$ locally by the graph
\[ \xi_k < f(\xi_1, \ldots, \xi_{k-1} \big) \:. \]
As in the proof of Lemma~\ref{lemmaeuclid}, we first consider the case~$f=0$
where~$\partial \Omega$ is a hyperplane. In this case, we obtain
\[g_\zeta(B, \nu) = 2 \delta^{k+1} \int_{T_{\zeta} N} |\xi_k| \: \eta \big( \|\xi\| \big)\:
\:\Big(1- \frac{\delta^2}{6} \Ric (\xi, \xi) + \mathcal{O} \big( (\delta\, |\xi|)^3 \big) \Big)\: d^k\xi \:, \]
and a straightforward computation gives the first term in~\eqref{gnu0}
as well as the terms in~\eqref{gnu}, with an error of order~$\O(\delta^{k+4})$.
Next, choosing~$f$ as a homogeneous polynomial of degree two,
a direct computation gives the last terms in~\eqref{gnu0}.
The error terms are obtained just as in the proof of Lemma~\ref{lemmaeuclid}.
\QED
Theorem~\ref{thmriemann} follows immediately by combining the results
of Lemmas~\ref{lemmariemann} and~\ref{lemma2}.

\section{General Notions of Dimension and Synthetic Scalar Curvature} \label{secscalar}
We choose a base point~$x_0 \in N$. According to our assumptions on the Lagrangian (see~(ii) and~(vi) in Section~\ref{secprelim}), the set
\[ 
K := \supp \L(x_0, \cdot ) \quad \text{is compact}\:. \]
We introduce the constant
\[ 
c_x := \sup_{x \in K, \;y \in N} \L(x,y) < \infty \]
(note that this constant is finite because~$\L$ is continuous and has compact range;
see assumptions~(ii) and~(vi) in Section~\ref{secprelim}).
We want to apply Theorem~\ref{thmlip} in case~(ii) for~$b$ a constant function
and the initial measure~$\nu=0$. Moreover, we choose a final volume~$V_{\max}>0$ with~$V_{\max} < \mu(N)$.
In order to satisfy the condition~\eqref{exvolcond2}, we need to choose the
external potential~$\phi$ as a function which is negative on a set of measure at least~$V_{\max}$.
To this end, it is useful to take the $k$-fold product of the Lagrangian,
\[ \ell_k(x) := \int_N d\mu(z_1) \cdots \int_N d\mu(z_k) \:\L(x_0, z_1)\: \L(z_1, z_2) \cdots \L(z_k, x) \:. \]
We choose~$k \in \N$ minimal such that
\[ (\mu - \nu) \Big( \big\{ x \in N \:\big|\: \ell_k(x) > 0 \big\} \Big) > V_{\max}\:. \]
Next, we choose the external potential as
\beq \label{phichoice}
\phi(x) = -\frac{s+2 b c_x}{\L(x_0, x_0)} \, \L(x, x_0) - \varepsilon\, \ell_k(x)
\eeq
with a parameter~$\varepsilon>0$. Finally, we choose~$b>0$ and set~$T=V_{\max}/b$. 

Applying Theorem~\ref{thmlip} in case~(ii), we obtain a
family of Lipschitz-continuous measures~$\rho(t)$ with~$t \in [0, T_{\max}]$ with~$\rho(0)=0$
and
\[ V \big( \rho(T_{\max}) \big) = b\, T_{\max} = V_{\max} \:. \]
In this way, we have constructed a measure of prescribed volume, which can be thought of as a
small {\em{isoperimetric ball}} which emanated from the single point~$x_0$. For ease in notation, we set
\beq \label{rhoV}
\rho := \rho(T_{\max}) \qquad \text{and} \qquad V := V_{\max} \:.
\eeq
Note that this construction still depends on the choice of the parameters~$\varepsilon, b$.
The parameter~$\varepsilon$ is needed for the existence of the flow. It seems a good
idea to take the limit inferior~$\varepsilon \searrow 0$ at the very end.
In simple terms, the parameter~$b$ determines how fast the volume increases in the flow.
For studying geometric properties, we are interested in the limiting case~$b \searrow 0$, where
the volume increases very slowly, so that the balls can settle down to isoperimetric balls
along the flow.
Using a notion from thermodynamics, we refer to this procedure that we increase the volume {\em{adiabatically}}. We expect that the adiabatic limit of the measure~$\rho(T_{\max})$
exists. But we do not want to prove it here. For our purposes, it suffices to consider the
flow for any~$b>0$ and take the limit inferior~$b \searrow 0$ at the end.

The next lemma shows that the isoperimetric balls contain the base point. This is an important observation,
because otherwise the isoperimetric ball would not encode information on the geometry near the base point.
\begin{Lemma} Assume that
\beq \label{Vmin}
V > V_{\min} := \mu(K_x) \:.
\eeq
Then the base point lies in the occupied set, i.e.\
\[ x_0 \in N_O \:. \]
\end{Lemma}
\Proof The assumption~\eqref{Vmin} ensures that there is a point~$x \in N_O$ with~$\L(x,x_0)=0$.
This makes it possible to estimate the parameter~$a$ in~\eqref{aval} by
\begin{align*}
a &\leq \int_N \L(\x,\y)\: d\nu(\y) - \phi(x) - \frac{1}{h}\:\epsilon(\x) \int_N \L(\x,\y)\: d|\rho-\nu|(\y) \\
&\leq \s + \varepsilon \ell_k(x) + \frac{1}{h}\: b\,h\: \sup_{y \in N} \L(x,y) \leq \s + \varepsilon \ell_k(x)+ b c_x \:,
\end{align*}
where in the last step we use the explicit form of~$\phi$ in~\eqref{phichoice}.
Using this estimate in~\eqref{elldef}, we obtain
\begin{align*}
\ell_\rho(x_0) &< \phi(x_0)+ a + \frac{1}{h}\:\int_N \L(x_0,\y)\: d|\rho-\nu|(\y) \\
&\leq \phi(x_0) + \s + \varepsilon \ell_k(x)+ 2 b c_x \:,
\end{align*}
which is negative for sufficiently small~$\varepsilon$ by our choice of the external potential~\eqref{phichoice}. Using~\eqref{ELnc}, we
conclude that~$x_0 \in N_O$.
\QED

Having the isoperimetric balls of volume~$V$ to our disposal, we can now
introduce general notions of local dimension and scalar curvature.
To this end, we need to specify a {\em{length scale}}. In order to explain the idea,
suppose that our space is a three-dimensional lattice of lattice spacing~$d$. Then, looked at on
the length scale~$d$, one sees the individual lattice points. Thus we have a zero-dimensional
object, and notions like scalar curvature do not seem to be sensible.
However, if we look at the lattice on a length scale which is much larger than the lattice spacing,
we may want to disregard the discrete microstructure. Then we have a three-dimensional space,
and geometric notions like scalar curvature do make sense.
Our notions are general enough to describe such situations. This also explains why we need
to specify a length scale on which the notions of dimension and scalar curvature are to be defined.
In order to specify the length scale, it is most convenient to use the volume of the isoperimetric balls.
We denote the corresponding parameter by~$V_\text{scale}$ and assume that it lies in the range
\[ V_{\min} < V_\text{scale} < \frac{\mu(N)}{2} \:. \]

In preparation, we recall how the dimension and scalar curvature enter the area and volume
of small geodesic balls on a Riemannian manifold. 
In $n$-dimensional Euclidean space, the area and volume a ball of radius~$r$ are given by
(see~\cite[eq.~5.19.4]{DLMF})
\[ V=\frac{\pi^{\frac{n}{2}}}{\Gamma\big(\frac{n}{2}+1\big)}\:\:r^{n}
\:,\qquad A =\frac{n}{r}\:V \]
Thus
\[ r = \frac{\Gamma\big(\frac{1}{2}n+1\big)^\frac{1}{n}}{\sqrt{\pi}}\: V^\frac{1}{n} \:, \]
making it possible to express the area as a function of volume by
\[ A =  \frac{n\,\sqrt{\pi}}{\Gamma\big(\frac{n}{2}+1\big)^\frac{1}{n}}\: V^\frac{n-1}{n}\:. \]
On a Riemannian manifold, scalar curvature changes the area and volume 
of small geodesic balls of radius~$r$ according to (see~\cite[Section~XII.8]{chavel})
\begin{align}
\frac{\operatorname{Vol}(B_r(p) \subset M)}{\operatorname{Vol}\left(B_r(0)\subset {\mathbb R}^n\right)}
&=1 - \frac{\text{scal}}{6(n + 2)}\: r^2 + O\left(r^3\right) \label{Vdefect} \\
\frac{\operatorname{Area} (\partial B_r(p) \subset M)}{\operatorname{Area}(\partial B_r(0)\subset {\mathbb R}^n)}
&= 1 - \frac{\text{scal}}{6n} \:r^2 + O\left(r^3\right) . \label{Adefect}
\end{align}
Combining these formulas, we obtain on a Riemannian manifold of dimension~$n$ that
for small geodesic balls, the area and volume are related by
\begin{align*}
A &=  \frac{n\,\sqrt{\pi}}{\Gamma\big(\frac{n}{2}+1\big)^\frac{1}{n}}\: V^\frac{n-1}{n} \:\Big( 1 - \frac{\text{scal}}{2n(n+2)} \: r^2+ \O \big( r^3 \big) \Big) \\
&= \frac{n\,\sqrt{\pi}}{\Gamma\big(\frac{n}{2}+1\big)^\frac{1}{n}}\: V^\frac{n-1}{n} 
- \frac{n\,\sqrt{\pi}}{\Gamma\big(\frac{n}{2}+1\big)^\frac{1}{n}}\: V^\frac{n-1}{n} \:
\frac{\text{scal}}{2n(n+2)}\: \frac{\Gamma\big(\frac{1}{2}n+1\big)^\frac{2}{n}}{\pi}\: V^\frac{2}{n}
+ \O \big( V^\frac{n+3}{n} \big) \:.
\end{align*}
We conclude that the area of small geodesic balls can be expressed in terms of their volume by
\[ 
A(V) = \frac{n\,\sqrt{\pi}}{\Gamma\big(\frac{n}{2}+1\big)^\frac{1}{n}}\: V^\frac{n-1}{n} 
- \frac{\Gamma\big(\frac{n}{2}+1\big)^\frac{1}{n}}{\sqrt{\pi}}
\frac{\text{scal}}{2(n+2)}\:  V^\frac{n+1}{n}
+ \O \big( V^\frac{n+3}{n} \big) \:. \]

This formula can be used in our setting to define dimension and scalar curvature.
To this end, we denote the area of the measure~$\rho$ in~\eqref{rhoV} describing the
isoperimetric ball of volume~$V$ by~$A(V)$. We now consider the areas~$A(V_\text{scale})$
and~$A(2 V_\text{scale})$. We define the {\em{local dimension at~$x$ on the scale $V_\text{scale}$}} by
\[ n \big(x, V_\text{scale} \big) := \frac{1}{1-e} \quad \text{with} \quad
e := \liminf_{\varepsilon \searrow 0} \liminf_{b \searrow 0} \frac{\log A(2 V_\text{scale}) - \log A(V_\text{scale})}{\log 2} \:. \]
Having determined~$n$, we define {\em{scalar curvature at~$x$ on the scale $V_\text{\rm{scale}}$}} by
\begin{align*}
&\text{scal} \big( x, V_\text{scale} \big) \\
&\;\;\!\!:= \liminf_{\varepsilon \searrow 0} \liminf_{b \searrow 0} \bigg( A(V_\text{scale}) 
- \frac{n\,\sqrt{\pi}}{\Gamma\big(\frac{n}{2}+1\big)^\frac{1}{n}}\: V_\text{scale}^\frac{n-1}{n} \bigg)
\bigg( - \frac{\sqrt{\pi}}{\Gamma\big(\frac{n}{2}+1\big)^\frac{1}{n}}
\frac{2(n+2)}{V_\text{scale}^\frac{n+1}{n}} \bigg) \\
&\;\;= \liminf_{\varepsilon \searrow 0} \liminf_{b \searrow 0} \bigg( \frac{2 n(n+2)\,\pi}{\Gamma\big(\frac{n}{2}+1\big)^\frac{2}{n}}
\frac{1}{V_\text{scale}^\frac{2}{n}}
- \frac{2(n+2)\,\sqrt{\pi}}{\Gamma\big(\frac{n}{2}+1\big)^\frac{1}{n}}\:
\frac{A(V_\text{scale})}{V_\text{scale}^\frac{n+1}{n}} \bigg) \:.
\end{align*}

We finally remark that the last formula could be made more robust by taking more than two
isoperimetric balls, or by forming an expression which is less sensitive to the value of~$n$. For example, one
could consider instead of~$A(V_\text{scale})$ the quantity
\[ \frac{\log A \big(2V_\text{scale} \big)}{\log (2 V_\text{scale})} - \frac{\log A \big(V_\text{scale} \big)}{\log V_\text{scale}} \:, \]
having the advantage that the leading contribution in~$n$ drops out.
Here we shall not work out resulting formulas in detail. The main point of our analysis to
show that isoperimetric balls can indeed be used to introduce notions of dimension and scalar curvature
in a fully general non-smooth setting.

\section{Comparison with Synthetic Scalar Curvature Coming from Quasi-Local Mass}
\label{secquasi}
In~\cite[Definition~9.1]{matter} a definition of synthetic scalar curvature was given in the context
of  a notion of a positive quasi-local mass.
In this section, we compare this notion with the definition given in Section~\ref{secscalar}.
To this end, we need to specialize our setting by assuming that curvature can be
described by a linear perturbation. Thus we assume that the measure~$\mu$ describes the vacuum.
By this we mean that the space~$N := \supp \mu$ can be identified with the Euclidean space~$\R^n$,
and the Lagrangian~$\L(x,y)$ depends only on the Euclidean distance~$\|x-y\|$.
Moreover, we assume that the interacting measure, denoted by~$\tilde{\mu}$, is a linear perturbation
of~$\mu$ described by the jet~$\v = (b, v)$.
For simplicity, we disregard the external potential~$\phi$.
Then the unperturbed isoperimetric flow simply is the flow of balls~$B_r :=B_r(0)$ with~$r>0$
The fact that each ball is isoperimetric means that for all~$x \in \partial B_r$,
\[ 
\int_{B_r} \L(x,y)\: d\mu(y) = a(\|x\|) := \frac{1}{2}\: \big(s - H(x) \big) \:, \]
with the mean curvature (defined by~\eqref{Hdef} with~$\rho := \chi_{B_r}\, \mu$) given by
\[ H(x) = \frac{\s_1\, \delta}{\|x\|} + \O\Big( \frac{\delta^2}{\|x\|^2} \Big)\:. \]
By a suitable parametrization and by adding a suitable inner solution, we can arrange that the perturbation
does not change the ball and leaves the Lagrange parameter~$a(\|x\|)$ unchanged, i.e.\
\[ \int_{B_{\|x\|}} \big(\nabla_{1,\v} + \nabla_{2,\v} \big) \L(x,y)\: d\mu(y) = b(x)\: a(\|x\|) \:. \]

Given~$x \in \tilde{N}$, in~\cite{matter} the vacuum~$N$
is transformed by a symmetry~$\Phi$ of the Lagrangian (see~\cite[Definition~6.1]{matter}),
with the goal of minimizing the quasi-local mass of a neighborhood of~$x$.
Using a notion introduced in~\cite{lcalc, gauss}, we refer to the transformed vacuum
as the {\em{osculating vacuum}}, denoted by~$M_x$. In our linearized description,
the symmetry of the Lagrangian is described by a vector field~$u$ on~$\G$ with the property that
\[ \big( D_{1,u} + D_{2,u} \big) \L(z,z') = 0 \qquad \text{for all~$z,z' \in N$}\:. \]
Transforming to the osculating vacuum amounts to arranging by a suitable choice of~$u$ that
\[ 
D_{1,v} \L(x,y) = 0 \:. \]
Next, one needs to specify the kernel~$\big(\nabla_{1,\v} + \nabla_{2,\v} \big) \L(x,y)$.
In~\cite{matter}, this problem is addressed under the assumption that, in a Gaussian coordinate
system around a point~$x$, the Lagrangian~$\L(x,.)$ should agree with that of Euclidean space
up to specified error term.
In the linearized description, this concept gives rise to the diffeomorphism term
and {\em{diffeomorphism jets}} (see~\cite[Appendix~B]{pmt} and~\cite[Section~10.3]{matter}).
Working with diffeomorphism jets, it was shown in~\cite[Theorem~10.6]{matter} that
in an ultrastatic spacetime~$\scrM = \R \times \scrN$, the synthetic scalar curvature agrees, up to a prefactor, with the scalar curvature of the Riemannian manifold~$(\scrN, g)$.
Therefore, using the results in Section~\ref{secriemann}, this notion of scalar curvature
also agrees with the notions introduced here.

There is an alternative, more direct method to get a connection between the
different definitions of scalar curvature. It also has the advantage that it suffices to assume
that~$\L(x,y)$ agrees with the Lagrangian of Euclidean space for~$y$ in a small
neighborhood of~$x$ (again with a specified error term).
To this end,
we choose a Gaussian coordinate system around~$x$. Then, using the assumption
that the Lagrangian agrees with that in Euclidean space, we know that
\beq \label{gaussian}
\big( D_{1,v} + D_{2,v} \big) \L(x,y) = 0
\eeq
(up to an error term which vanishes in the continuum limit and which, for
brevity, will be disregarded here).

Denoting the corresponding jet by~$\v=(b,v)$, the linearized field equations imply that
\[ \int_N \big( \nabla_{1,\v} + \nabla_{2,\v} \big) \L(x,y)\: d\mu(y) - \nabla_{1,\v} \s = 0 \:. \]
Using the EL equations together with~\eqref{gaussian}, this simplifies to
\beq \label{bdet}
0 = \int_N \big( D_{1,\v} + \nabla_{2,\v} \big) \L(x,y)\: d\mu(y)
= \int_N b(y)\: \L(x,y)\: d\mu(y) \:.
\eeq
Now we make use of the fact that, in our Gaussian coordinates, the volume measure~$\tilde{\mu}$
differs from~$\mu$ by a term quadratic in the distance and linear in scalar curvature
(see~\eqref{Vdefect}). Therefore,
\[ \tilde{\mu} = h\: \mu \qquad \text{with} \qquad
h(y) = 1 - c\: \text{scal}(x_0)\: d(x_0, y)^2\:. \]
The proportionality factor can be determined from~\eqref{bdet}. We conclude that
\[ b(x) = \big( c_1 \delta^2 - c_2 r^2 \big)\:\text{scal}(x) \:. \]
Finally, the scalar curvature introduced in~\cite[Definition~9.1]{matter} is given by
\[ 
\nabla_\v \s - \Delta \v = b(x)\, \s =  c_1 \delta^2\, \s \:\text{scal}(x) \:. \]
This shows that the notions of scalar curvature indeed agree, up to a prefactor
which involves the range of the Lagrangian.

In this way, one sees that the synthetic scalar curvature as introduced in Section~\ref{secscalar}
harmonizes with the notion introduced previously in~\cite{matter}.
The notion introduced here has the advantage that it is much more general.
In particular, we do not need to assume that~$N$ has a smooth manifold structure.
Moreover, we have the freedom to choose the length scale on which
scalar curvature should be defined. In the notion in~\cite{matter}, however,
this length scale necessarily coincides with the range~$\delta$ of the Lagrangian.
This is too restrictive for instance in situations when space or spacetime has a non-smooth
or even discrete structure on the scale~$\delta$. \\


\bibliographystyle{amsplain}

\begin{thebibliography}{10}

\bibitem{cfsweblink}
\emph{Link to web platform on causal fermion systems:
  \href{https://www.causal-fermion-system.com}{\textrm{www.causal-fermion-system.com}}}.

\bibitem{ambrosio-minmov}
L.~Ambrosio, \emph{Minimizing movements}, Rend. Accad. Naz. Sci. XL Mem. Mat.
  Appl. (5) \textbf{19} (1995), 191--246.

\bibitem{ambrosio+gigli}
L.~Ambrosio, N.~Gigli, and G.~Savar\'{e}, \emph{Gradient flows in metric spaces
  and in the space of probability measures}, second ed., Lectures in
  Mathematics ETH Z\"{u}rich, Birkh\"{a}user Verlag, Basel, 2008.

\bibitem{AGS-RCD}
L.~Ambrosio, N.~Gigli, and G.~Savar\'e, \emph{Metric measure spaces with
  {R}iemannian {R}icci curvature bounded from below}, Duke Math. J.
  \textbf{163} (2014), no.~7, 1405--1490.

\bibitem{ambrosio+tilli}
L.~Ambrosio and P.~Tilli, \emph{Topics on {A}nalysis in {M}etric {S}paces},
  Oxford Lecture Series in Mathematics and its Applications, vol.~25, Oxford
  University Press, Oxford, 2004.

\bibitem{berger-gauduchon}
M.~Berger, P.~Gauduchon, and E.~Mazet, \emph{Le {S}pectre d'une {V}ari\'et\'e{}
  {R}iemannienne}, Lecture Notes in Mathematics, vol. Vol. 194,
  Springer-Verlag, Berlin-New York, 1971.

\bibitem{BBM}
J.~Bourgain, H.~Brezis, and P.~Mironescu, \emph{Another look at {S}obolev
  spaces}, Optimal control and partial differential equations, IOS, Amsterdam,
  2001, pp.~439--455.

\bibitem{braides-minmov}
A.~Braides, \emph{Local {M}inimization, {V}ariational {E}volution and
  {$\Gamma$}-{C}onvergence}, Lecture Notes in Mathematics, vol. 2094, Springer,
  Cham, 2014.

\bibitem{CRS}
L.~Caffarelli, J.-M. Roquejoffre, and O.~Savin, \emph{Nonlocal minimal
  surfaces}, \href{https://arxiv.org/abs/0905.1183}{arXiv:0905.1183 [math.AP]},
  Comm. Pure Appl. Math. \textbf{63} (2010), no.~9, 1111--1144.

\bibitem{chavel}
I.~Chavel, \emph{Eigenvalues in {R}iemannian {G}eometry}, Pure and Applied
  Mathematics, vol. 115, Academic Press, Inc., Orlando, FL, 1984.

\bibitem{degiorgi-minmov}
E.~De~Giorgi, \emph{New problems on minimizing movements}, Boundary value
  problems for partial differential equations and applications, RMA Res. Notes
  Appl. Math., vol.~29, Masson, Paris, 1993, pp.~81--98.

\bibitem{docarmo-riemannian}
M.P. do~Carmo, \emph{Riemannian {G}eometry}, Mathematics: Theory \&
  Applications, Birkh\"auser Boston, Inc., Boston, MA, 1992.

\bibitem{continuum}
F.~Finster, \emph{Causal variational principles on measure spaces},
  \href{https://arxiv.org/abs/0811.2666}{arXiv:0811.2666 [math-ph]}, J. Reine
  Angew. Math. \textbf{646} (2010), 141--194.

\bibitem{cfs}
\bysame, \emph{The {C}ontinuum {L}imit of {C}ausal {F}ermion {S}ystems},
  \href{https://arxiv.org/abs/1605.04742}{arXiv:1605.04742 [math-ph]},
  Fundamental Theories of Physics, vol. 186, Springer, Cham, 2016.

\bibitem{nonlocal}
\bysame, \emph{Solving the linearized field equations of the causal action
  principle in {M}inkowski space},
  \href{https://arxiv.org/abs/2304.00965}{arXiv:2304.00965 [math-ph]}, Adv.
  Theor. Math. Phys. \textbf{27} (2023), no.~7, 2087--2217.

\bibitem{minmov}
F.~Finster and F.~Gmeineder, \emph{Action-driven flows for causal variational
  principles}, \href{https://arxiv.org/abs/2503.00526}{arXiv:2503.00526
  [math-ph]}, Forum Mathematics, Sigma \textbf{14} (2026), e83.

\bibitem{cauchy}
F.~Finster and A.~Grotz, \emph{On the initial value problem for causal
  variational principles},
  \href{https://arxiv.org/abs/1303.2964}{arXiv:1303.2964 [math-ph]}, J. Reine
  Angew. Math. \textbf{725} (2017), 115--141.

\bibitem{review}
F.~Finster and M.~Jokel, \emph{Causal fermion systems: {A}n elementary
  introduction to physical ideas and mathematical concepts},
  \href{https://arxiv.org/abs/1908.08451}{arXiv:1908.08451 [math-ph]},
  {P}rogress and {V}isions in {Q}uantum {T}heory in {V}iew of {G}ravity
  (F.~Finster, D.~Giulini, J.~Kleiner, and J.~Tolksdorf, eds.), Birkh\"auser
  Verlag, Basel, 2020, pp.~63--92.

\bibitem{matter}
F.~Finster and N.~Kamran, \emph{A positive quasilocal mass for causal
  variational principles},
  \href{https://arxiv.org/abs/2310.07544}{arXiv:2310.07544 [math-ph]}, Calc.
  Var. Partial Differential Equations \textbf{64} (2025), no.~3, 91.

\bibitem{lcalc}
F.~Finster, N.~Kamran, and F.~van~der Top, \emph{The {$\mathcal{L}$}-calculus
  for causal variational principles: {A}n exterior differential calculus on
  non-smooth spaces}, \href{https://arxiv.org/abs/2608.08811}{arXiv:2608.08811
  [math-ph]} (2026).

\bibitem{intro}
F.~Finster, S.~Kindermann, and J.-H. Treude, \emph{{C}ausal {F}ermion
  {S}ystems: {A}n {I}ntroduction to {F}undamental {S}tructures, {M}ethods and
  {A}pplications}, \href{https://arxiv.org/abs/2411.06450}{arXiv:2411.06450
  [math-ph]}, Cambridge Monographs on Mathematical Physics, Cambridge
  University Press, Cambridge, 2026.

\bibitem{dice2014}
F.~Finster and J.~Kleiner, \emph{Causal fermion systems as a candidate for a
  unified physical theory},
  \href{https://arxiv.org/abs/1502.03587}{arXiv:1502.03587 [math-ph]}, J.
  Phys.: Conf. Ser. \textbf{626} (2015), 012020.

\bibitem{jet}
\bysame, \emph{A {H}amiltonian formulation of causal variational principles},
  \href{https://arxiv.org/abs/1612.07192}{arXiv:1612.07192 [math-ph]}, Calc.
  Var. Partial Differential Equations \textbf{56:73} (2017), no.~3, 33.

\bibitem{gauss}
F.~Finster and C.~Krpoun, \emph{A geometric derivation of the {E}instein
  equations from the causal action principle},
  \href{https://arxiv.org/abs/2607.13871}{arXiv:2607.13871 [math-ph]} (2026).

\bibitem{noncompact}
F.~Finster and C.~Langer, \emph{Causal variational principles in the
  $\sigma$-locally compact setting: {E}xistence of minimizers},
  \href{https://arxiv.org/abs/2002.04412}{arXiv:2002.04412 [math-ph]}, Adv.
  Calc. Var. \textbf{15} (2022), no.~3, 551--575.

\bibitem{pmt}
F.~Finster and A.~Platzer, \emph{A positive mass theorem for static causal
  fermion systems}, \href{https://arxiv.org/abs/1912.12995}{arXiv:1912.12995
  [math-ph]}, Adv. Theor. Math. Phys. \textbf{25} (2021), no.~7, 1735--1818.

\bibitem{gigli-nonsmooth}
N.~Gigli, \emph{Nonsmooth differential geometry --- an approach tailored for
  spaces with {R}icci curvature bounded from below},
  \href{https://arxiv.org/abs/1407.0809}{arXiv:1407.0809 [math.DG]}, Mem. Amer.
  Math. Soc. \textbf{251} (2018), no.~1196, v+161.

\bibitem{gray}
A.~Gray, \emph{Tubes}, second ed., Progress in Mathematics, vol. 221,
  Birkh\"auser Verlag, Basel, 2004.

\bibitem{lott-villani}
J.~Lott and C.~Villani, \emph{Ricci curvature for metric-measure spaces via
  optimal transport},
  \href{https://arxiv.org/abs/math/0412127}{arXiv:math/0412127 [math.DG]}, Ann.
  of Math. (2) \textbf{169} (2009), no.~3, 903--991.

\bibitem{mantegazza}
C.~Mantegazza, \emph{Lecture {N}otes on {M}ean {C}urvature {F}low}, Progress in
  Mathematics, vol. 290, Birkh\"auser/Springer Basel AG, Basel, 2011.

\bibitem{miranda}
M.~Miranda, Jr., \emph{Functions of bounded variation on ``good'' metric
  spaces}, J. Math. Pures Appl. (9) \textbf{82} (2003), no.~8, 975--1004.

\bibitem{DLMF}
F.W.J. Olver, D.W. Lozier, R.F. Boisvert, and C.W. Clark (eds.), \emph{Digital
  {L}ibrary of {M}athematical {F}unctions}, National Institute of Standards and
  Technology from \href{http://dlmf.nist.gov/}{http://dlmf.nist.gov/} (release
  date 2019-12-15), Washington, DC, 2010.

\bibitem{ponce}
A.C. Ponce, \emph{A new approach to {S}obolev spaces and connections to
  {$\Gamma$}-convergence}, Calc. Var. Partial Differential Equations
  \textbf{19} (2004), no.~3, 229--255.

\bibitem{ritore-sinestrari}
M.~Ritor\'e and C.~Sinestrari, \emph{Mean {C}urvature {F}low and
  {I}soperimetric {I}nequalities}, Advanced Courses in Mathematics. CRM
  Barcelona, Birkh\"auser Verlag, Basel, 2010.

\bibitem{sasaki}
S.~Sasaki, \emph{On the differential geometry of tangent bundles of
  {R}iemannian manifolds}, T\^ohoku Math. J. (2) \textbf{10} (1958), 338--354.

\bibitem{shioya}
T.~Shioya, \emph{Metric {M}easure {G}eometry},
  \href{https://arxiv.org/abs/1410.0428}{arXiv:1410.0428 [math.MG]}, IRMA
  Lectures in Mathematics and Theoretical Physics, vol.~25, EMS Publishing
  House, Z\"urich, 2016.

\bibitem{sturm-mms}
K.-T. Sturm, \emph{Metric measure spaces and synthetic {R}icci bounds:
  fundamental concepts and recent developments},
  \href{https://arxiv.org/abs/2404.15755}{arXiv:2404.15755 [math.MG]}, European
  {C}ongress of {M}athematics, EMS Press, Berlin, 2023, pp.~125--159.

\bibitem{sturmI}
K.-Th. Sturm, \emph{On the geometry of metric measure spaces. {I}}, Acta Math.
  \textbf{196} (2006), 65--131.

\bibitem{sturmII}
\bysame, \emph{On the geometry of metric measure spaces. {II}}, Acta Math.
  \textbf{196} (2006), 133--177.

\end{thebibliography}
\providecommand{\bysame}{\leavevmode\hbox to3em{\hrulefill}\thinspace}
\providecommand{\MR}{\relax\ifhmode\unskip\space\fi MR }
\providecommand{\MRhref}[2]{%
  \href{http://www.ams.org/mathscinet-getitem?mr=#1}{#2}
}
\providecommand{\href}[2]{#2}

\end{document}